\documentclass[onecolumn]{svjour3}
\smartqed  

\usepackage[numbers,comma,sort,compress]{natbib}
\usepackage{setspace}
\usepackage{mathtools}
\usepackage{amsfonts,amssymb,amsbsy,amsmath}
\usepackage[font=small,labelfont=bf]{caption}
\usepackage{subcaption}
\usepackage{graphicx}
\usepackage{xifthen}

\usepackage{bigints}
\usepackage{booktabs}
\usepackage{nicefrac}
\usepackage{color}
\usepackage{url}
\usepackage[usenames,dvipsnames,table]{xcolor}
\usepackage{fancyhdr}
\usepackage{perpage}
\usepackage{hyperref} 
\usepackage{setspace} 
\usepackage[version=4]{mhchem} 
\usepackage[theorems]{tcolorbox}
\usepackage{enumitem}
\usepackage{boxedminipage}
\usepackage{wrapfig}
\usepackage[labelfont=bf]{caption}
\usepackage[a4paper, margin=1in]{geometry}
\usepackage{gensymb} 
\usepackage{tabularx}
\usepackage{array} 
\usepackage{import} 
\usepackage{pifont,latexsym,ifthen,theorem,rotating,calc}
\usepackage{multirow}

\theoremstyle{remark}

\usepackage{array}

\makeatletter
\newcommand{\thickhline}{%
    \noalign {\ifnum 0=`}\fi \hrule height 2pt
    \futurelet \reserved@a \@xhline
}
\catcode`_=\active
\newcommand_[1]{\ensuremath{\sb{\mathrm{#1}}}}

\catcode`^=\active
\newcommand^[1]{\ensuremath{\sp{\mathrm{#1}}}}

\newcommand{\ceq}[1]{\mathrel{\stackrel{\makebox[0pt]{\mbox{\normalfont\tiny #1}}}{=}}}

\newcommand{\indic}{{\mathrm{\bs{1}}}}    

\newcommand{\eg}{{e.g.,}}
\newcommand{\ie}{{i.e.,}}
\newcommand{\iid}{{i.i.d.}}

\newcommand{\pdfbig}[1]{{\pi\big(#1\big)}}

\newcommand{\numChar}{{n}} 
\newcommand{\varChar}{{a}}
\newcommand{\obsChar}{{o}}
\newcommand{\paraChar}{{p}}
\newcommand{\dataChar}{{d}}
\newcommand{\physChar}{{p}}
\newcommand{\aleaChar}{{i}}
\newcommand{\episChar}{{n}}
\newcommand{\aleaPhys}{{\physChar\aleaChar}}

\newcommand{\ndv}{{\numChar_{\dataChar\varChar}}} 
\newcommand{\npp}{{\numChar_{\paraChar\physChar}}} 
\newcommand{\npa}{{\numChar_{\paraChar\aleaChar}}} 
\newcommand{\npe}[1][]{{ \numChar_{ \ifthenelse{\isempty{#1}}{\paraChar\episChar}{{\paraChar\episChar,#1}} } }} 
\newcommand{\ndo}{{\numChar_{\dataChar\obsChar}}} 

\newcommand{\phys}{{phys}}

\newcommand{\alea}{{inad}}
\newcommand{\epis}{{nois}}

\newcommand{\scen}{{\mathcal{S}}}
\newcommand{\pscen}{{\mathcal{S}_\phys}}

\def\gtrsim{\mathrel{\hbox{\rlap{\hbox{\lower4pt\hbox{$\sim$}}}\hbox{$>$}}}}
\def\lessim{\mathrel{\hbox{\rlap{\hbox{\lower4pt\hbox{$\sim$}}}\hbox{$<$}}}}

\newcommand{\up}{\operatorname}
\newcommand{\diff}{{\up{d}}}

\newcommand{\bs}{\boldsymbol}

\newcommand{\model}{{M}}
\newcommand{\pmodel}{{\bs\model_\phys}}
\newcommand{\amodel}{{\bs\model_\alea}}
\newcommand{\emodel}[1][]{{ \bs\model_{ \ifthenelse{\isempty{#1}}{\epis}{{\epis,#1}} } }}

\newcommand{\pamodel}{{\bs\model_\aleaPhys}}

\newcommand{\emodelset}{{\mathcal{M}_\epis}}
\newcommand{\eparamset}{{\Theta_\epis}}

\newcommand{\param}{{\bs{\theta}}}
\newcommand{\paramopt}{{\widehat\param}}
\newcommand{\pparam}{{\param_\phys}}
\newcommand{\aparam}{{\param_\alea}}
\newcommand{\eparam}[1][]{{ \param_{ \ifthenelse{\isempty{#1}}{\epis}{{\epis,#1}} } }}

\newcommand{\paparam}{{\param_\aleaPhys}}

\newcommand{\pparamopt}{{\paramopt_\phys}}

\newcommand{\pspace}{{\bs{\mathrm\Theta}}}
\newcommand{\ppspace}{{\pspace_\phys}}
\newcommand{\paspace}{{\pspace_\alea}}
\newcommand{\pespace}{{\pspace_\epis}}
\newcommand{\sspace}{{ \bs{\up{\Omega}}_{\mathrm\truth} }}
\newcommand{\psspace}{{ \bs{\up{\Omega}}_{\mathrm\ptruthset} }}
\newcommand{\ppaspace}{{\pspace_{\physChar\aleaChar}}}

\newcommand{\dataTriplet}{{ \{\dataset,\eparamset,\emodelset\} }}

\newcommand{\rand}{{\bs U}}
\newcommand{\prand}{{\rand^\possible}}
\newcommand{\randset}{{\mathcal{U}}}
\newcommand{\prandset}{{\randset^\possible}}

\newcommand{\tvar}{{R}} 
\newcommand{\dvar}{{D}} 
\newcommand{\data}{{\bs{\dvar}}}

\newcommand{\truth}{{\bs{\tvar}}}
\newcommand{\possible}{{*}}
\newcommand{\ptruth}{{\bs{\tvar}^\possible}} 
\newcommand{\truthset}{{\mathcal{\tvar}}}
\newcommand{\ptruthset}{{\truthset^\possible}}
\newcommand{\truthsubset}[1][]{{ \truthset_{ \ifthenelse{\isempty{#1}}{\truth}{{\truth_{#1}}} } }}
\newcommand{\ptruthsubset}[1][]{{ \truthset_{ \ifthenelse{\isempty{#1}}{\truth}{{\truth_{#1}}} }^\possible }}
\newcommand{\ptruthSuperset}{{\psspace}}
\newcommand{\ptruthSupersetPparam}{{ \ptruthSuperset\big(\pparam\big) }}
\newcommand{\dataset}{{\mathcal{\dvar}}}
\newcommand{\datasubset}[1][]{{ \dataset_{ \ifthenelse{\isempty{#1}}{\data}{{\data_{#1}}} } }}

\newcommand{\tvarind}[2][]{{ \tvar^{#2}_{{#1}} }}
\newcommand{\tvardep}[2][]{{ \tvar^{#2}_{{#1}} }}
\newcommand{\dvarind}[2][]{{ \dvar^{#2}_{{#1}} }}
\newcommand{\dvardep}[2][]{{ \dvar^{#2}_{{#1}} }}
\newcommand{\truthind}[1][]{{ \truth^{ind}_{ \ifthenelse{\isempty{#1}}{}{{#1}} } }}
\newcommand{\truthdep}[1][]{{ \truth^{dep}_{ \ifthenelse{\isempty{#1}}{}{{#1}} } }}
\newcommand{\dataind}[1][] {{  \data^{ind}_{ \ifthenelse{\isempty{#1}}{}{{#1}} } }}
\newcommand{\datadep}[1][] {{  \data^{dep}_{ \ifthenelse{\isempty{#1}}{}{{#1}} } }}

\newcommand{\prior}{{\mathcal{I}}}

\newcommand{\like}{{\mathcal{L}}}

\newcommand{\xx}[1][]{{ \ifthenelse{\isempty{#1}}{\textcolor{red}{XXX}}{\textcolor{red}{~(XXX {#1} XXX)~}} }} 

\newcolumntype{"}{@{\hskip\tabcolsep\vrule width 2pt\hskip\tabcolsep}}
\makeatother

\graphicspath{
{./figures/intro/}
{./figures/phys/}
{./figures/physInad/}
{./figures/physNois/}
{./figures/physInadNois/}
} 

\begin{document}
\onehalfspacing
\large
\title{\bf\Large Multilevel Bayesian Parameter Estimation in the Presence of Model Inadequacy and Data Uncertainty} 

\titlerunning{Parameter Estimation in the presence of Model Inadequacy and Data Uncertainty} 

\author{
       \normalfont
       Amir Shahmoradi~$^{1,2,3,\star}$ \\
       } \vspace{-0.3in}

\authorrunning{
                  A. Shahmoradi
              }

\institute{ \small
            $^1$ Center For Computational Oncology, The University of Texas at Austin, TX 78712 \\
            $^{2}$ Institute for Computational Engineering and Sciences, The University of Texas at Austin, TX 78712 \\
            $^{3}$ Department of Aerospace Engineering and Engineering Mechanics, The University of Texas at Austin, TX 78712 \\
            $^\star$ Peter O'Donnell, Jr. Fellow, \email{amir@physics.utexas.edu}
}

\date{\today}

\maketitle

\begin{abstract}
    Model inadequacy and measurement uncertainty are two of the most confounding aspects of inference and prediction in quantitative sciences. The process of scientific inference (the inverse problem) and prediction (the forward problem) involve multiple steps of data analysis, hypothesis formation, model construction, parameter estimation, model validation, and finally, the prediction of the quantity of interest. This article seeks to clarify the concepts of model inadequacy and bias, measurement uncertainty, and the two traditional classes of uncertainty: aleatoric versus epistemic, as well as their relationships with each other in the process of scientific inference. Starting from basic principles of probability, we build and explain a hierarchical Bayesian framework to quantitatively deal with model inadequacy and noise in data. The methodology can be readily applied to many common inference and prediction problems in science, engineering, and statistics.
\end{abstract}

\newpage

\tableofcontents

{\small
\begin{table}[htbp!]
\caption{Nomenclature and Definitions of Symbols in This Manuscript} \vspace{-3pt} \label{tab:symbols}
\begin{tabular}{p{0.08\textwidth} p{0.87\textwidth}}
    \toprule\toprule
    $\truth             $ & The reality or truth representing one event, without any observational bias or uncertainty. \\
    $\truthset          $ & The set containing the truth $\truth$ for each individual event that is observed. \\
    $\sspace            $ & The observational sampling space to which each event in the dataset belongs: $\truth\in\truthset\subset\sspace$. \\
    $\truth_i           $ & The reality or truth, $\truth$, about the $i$th event in event-set $\truthset$. \\
    $\truthsubset[i]    $ & A subset of $\truthset$ on which the $i$th event, $\truth_i$, depends. It can be and often is a null set. \\
    $\data              $ & All data about an event {\it as observed}, which is subject to measurement error, unlike $\truth$. \\
    $\dataset           $ & The set of all observations, each of which corresponds to one unique event, $\truth_i\in\truthset$. \\
    $\data_i            $ & All observational data, $\data$, about the $i$th observation in dataset $\dataset$. \\
    $\rand              $ & A stochastic variable representing difference between $\truth$ and the output, $\truth'$, of $\pmodel$. \\
    $\randset           $ & The set of all $\rand_i$, each of which corresponds to one event $\truth_i\in\truthset$. \\
    $\ptruth            $ & One possible realization of $\truth$, given $\data$ and the corresponding noise model, $\emodel$. \\
    $\ptruthset         $ & One possible realization of $\truthset$, given $\dataset$ and the set of noise models, $\emodelset$. \\
    $\ptruthsubset[i]   $ & One possible realization of $\truthsubset[i]$. \\
    $\ptruthSuperset    $ & The (super)set containing all possible realizations, $\ptruthset$, of the set $\truthset$. \\
    $\pmodel            $ & The physical model hypothesized to hold for the collection of events in $\truthset$. \\
    $\amodel            $ & The statistical physics-based model that quantifies the inadequacy of $\pmodel$ in describing $\truthset$. \\ 
    $\emodel            $ & The statistical model that quantifies the experimental measurement uncertainty (noise) in $\dataset$. \\
    $\emodelset         $ & The set of statistical models $\emodel[i]$ corresponding to each observation $\data_i\in\dataset$. \\
    $\pparam            $ & The vector of parameters of the physical model $\pmodel$. \\
    $\aparam            $ & The vector of parameters of the inadequacy model $\amodel$. \\
    $\eparam            $ & The vector of parameters of the noise model $\emodel$. \\
    $\pspace_{(\cdot)}  $ & The parameter space of a model, $(\cdot)$ -- \eg~$\pmodel: \ppspace$, $\amodel: \paspace$, $\emodel: \pespace$. \\
    $\ndo               $ & The number of data observations in $\dataset$ (or equivalently, the number of events in $\truthset$). \\
    $\ndv               $ & The number of data attributes by which each event is characterized; length of $\truth$ \& $\data$. \\
    $\npp               $ & The number of parameters of the physical model;  length of $\pparam$; dimension of $\ppspace$. \\
    $\npa               $ & The number of parameters of the inadequacy model; length of $\aparam$; dimension of $\paspace$. \\
    $\npe               $ & The number of parameters of the noise model; length of $\eparam$; dimension of $\pespace$. \\
    $\mathbb{R}         $ & The set of real numbers. \\
    $(\cdot)'           $ & The output of $\pmodel$, which may or may not be identical to the input entity, $(\cdot)$, to $\pmodel$. \\
    $\pi(\cdot)         $ & The Probability Density Function (PDF) of a statistical model. \\
    $\like(\cdot)       $ & The likelihood function of the parameters of a model. \\
    $\prior_{(\cdot)}   $ & The prior knowledge about the subscript entity, $(\cdot)$. \\
    \bottomrule
\end{tabular}
\end{table}
}

\newpage

\onehalfspacing

\section{Introduction}
\label{sec:intro}

\begin{figure}[t!]
    \centering
    \begin{subfigure}[t]{0.49\textwidth}
        \centering
        \includegraphics[width=\textwidth]{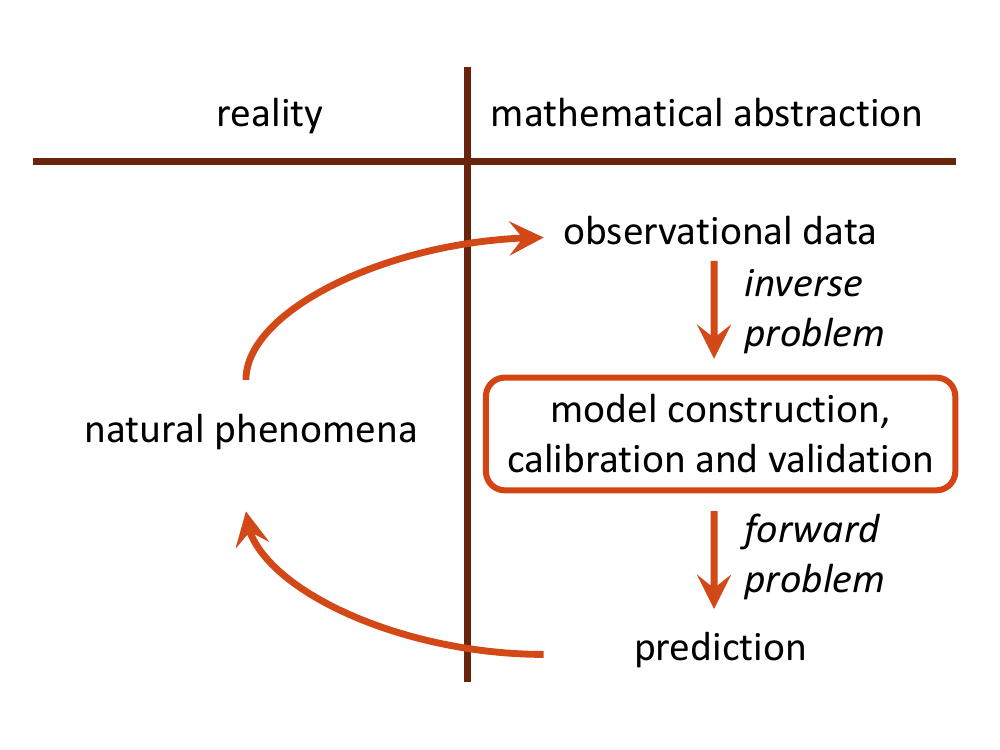}
        \caption{The scientific methodology.} \label{fig:scientificMethod}
    \end{subfigure}
    \begin{subfigure}[t]{0.49\textwidth}
        \centering
        \includegraphics[width=\textwidth]{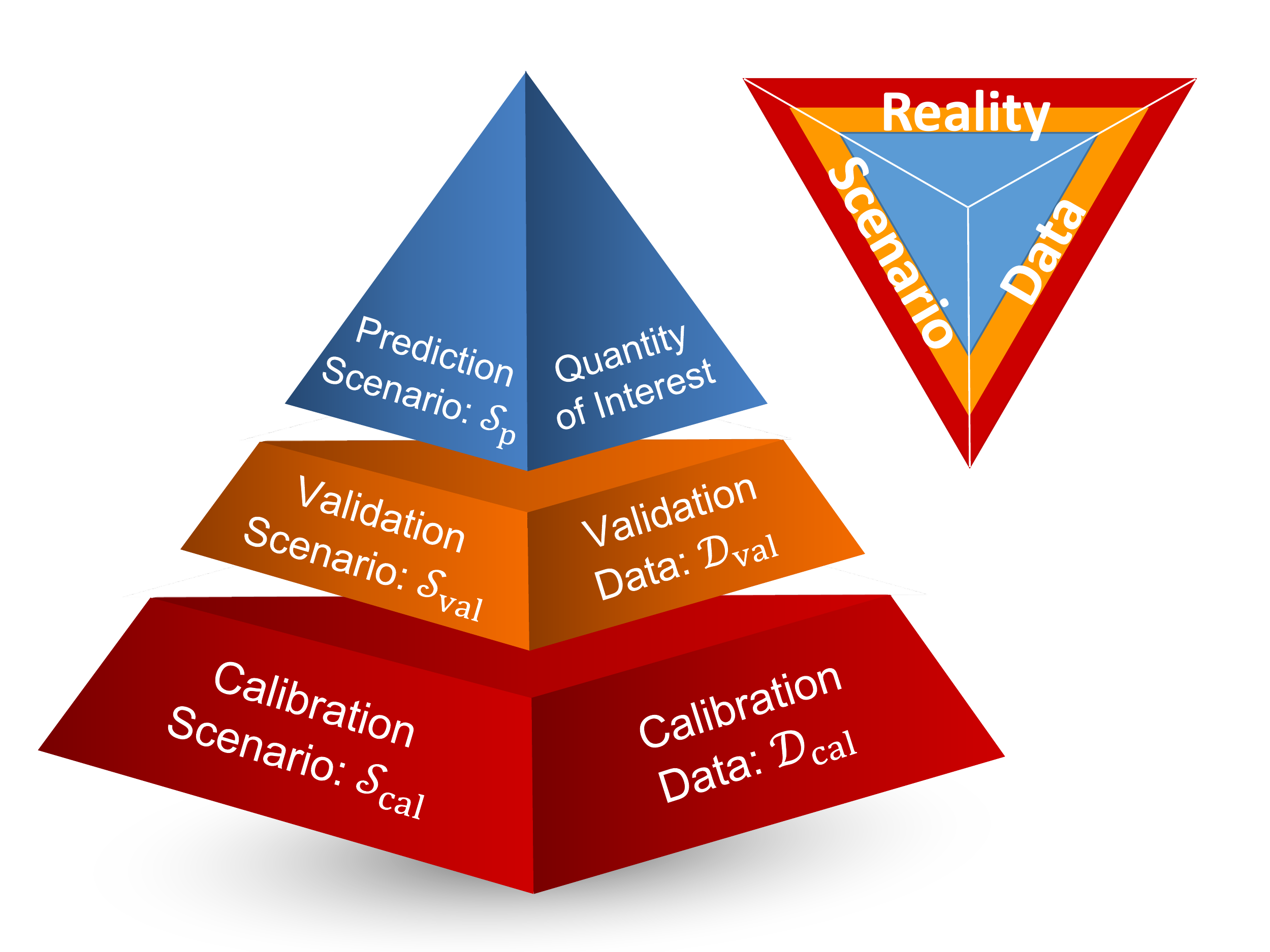}
        \caption{The prediction pyramid.} \label{fig:pyramid}
    \end{subfigure}
    \caption{{\bf(a)} {\bf The steps of scientific methodology}, involving data collection, hypothesis formulation, construction of a mathematical model and objective function, which is subsequently optimized to constrain the parameters of the model, a process known as {\it inversion} or {\it inverse problem}. Once validated, the model can be used to make predictions about the quantity of interest ({\it forward problem}). {\bf(b)} {\bf The prediction pyramid}, depicting the three hierarchical levels of predictive inference from bottom to top: Calibration, Validation, and Prediction of the Quantity of Interest (QoI). The rear face of the tetrahedron represents reality (truth), $\truthset$, about the set of observed phenomena, which is never known to the observer. The right-front face of the tetrahedron represents the observational data, $\dataset$, which results from the convolution of the truth/reality, $\truthset$, with various forms of measurement uncertainty. The left-front face represents the scenarios, $\scen$, under which data is collected, as well as the set of models that are hypothesized to describe the unknown truth, $\truthset$ \cite{oden2004predictive, oden2010computer, oden2013selection, oden2017}.}
\end{figure}

The process of scientific inference involves the collection of experimental data from observations of a set of natural phenomena, the analysis and reduction of the collected dataset, the formulation of a hypothesis (i.e., development of a physics-based mathematical model) that attempts to explain various potential causal relationships between different characteristics of data, and, finally, testing the predictions of the proposed model against new observational data by performing new experiments (Figure \ref{fig:scientificMethod}). \\

In the majority of scientific problems, the proposed physical model involves a set of parameters that have to be tuned in order to best describe the available data. For example, Einstein's famous equation of mass-energy equivalence, $E=mc^2$, relates the mass, $m$, of any material to an equivalent amount of energy, $E$, via an a-priori unknown constant, $c$, that is the speed of light, which has to be determined by experimental data. The process of inferring the parameters of the physical model is commonly known as {\it inversion} or an {\it inverse problem}, which could also be regarded as {\it model calibration} (Figure \ref{fig:pyramid}). \\

Once the parameters of a physical model are constrained, the proposed physical model has to be verified and its predictions validated against a new independent dataset. Extensive literature already exists on the topic of model verification and validation \cite[\eg~][]{roache1998verification, knupp2002verification, babuska2004verification, babuvska2005reliability, oberkampf2006measures, babuska2007reliability, babuvska2008systematic, oden2011control, farrell2015bayesian, oden2015estimation} as well as on decision theory \cite[for elegant reviews from a Bayesian perspective, see][]{jeffrey1992probability, jaynes2003probability, lindley2006understanding}. The validated model can be then used to make predictions of the Quantities of Interest (QoI), the precise physical features of the response of the system targeted in the simulation. This prediction step is commonly known as the {\it forward problem} in the scientific literature. \\

\begin{figure}[t!]
    \begin{center}
        \includegraphics[width=\textwidth]{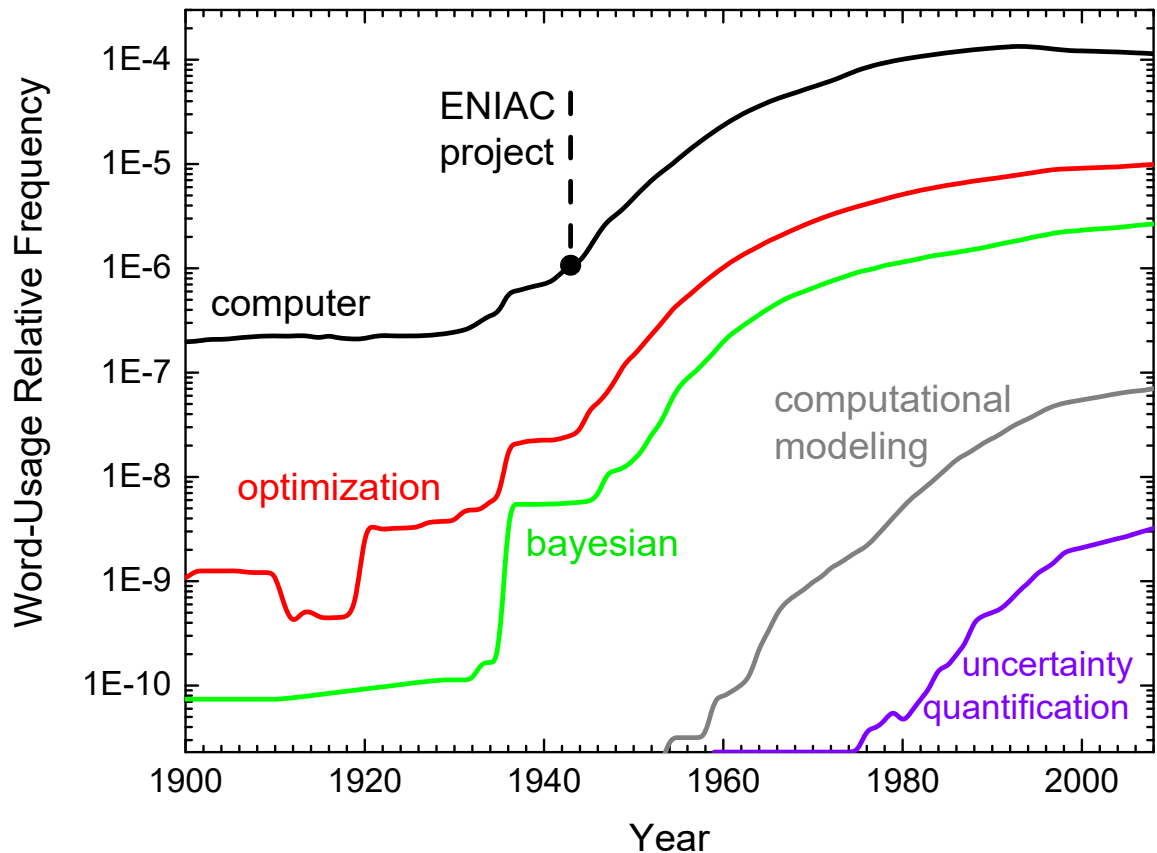}
    \end{center}
    \caption{A word-usage relative frequency plot \cite{michel2011quantitative}, illustrating the exponential growth of computer technology in the mid 20th century, as well as developments in the fields of deterministic and stochastic optimization techniques, which ultimately led to the emergence of `computational modeling' as the third pillar of science \citep{oden2010computer}. Advances in the computational methods and technology also led to the gradual popularity of Bayesian techniques in mathematical modeling toward the end of the 20th century, as well as the emergence of {\it uncertainty quantification} as a new field of science. Note that the positive-slope linear behavior on this semi-logarithmic plot implies exponential growth.}
    \label{fig:wordFreq}
\end{figure}

The process of scientific inference described above, although straightforward at a first glance, is severely complicated by the presence of many sources of uncertainty in multiple levels of data acquisition and model construction, as well as the inverse and forward problems. In fact, the significance of the effects of uncertainty in data and modeling has led to the emergence of a new field of science within the past three decades, specifically dedicated to {\it Uncertainty Quantification} (Figure \ref{fig:wordFreq}). \\

\subsection{Aleatoric vs. Epistemic Uncertainty}
\label{sec:introAleaEpis}

Different classes and sources of uncertainty have been already identified and extensively studied at different levels of the scientific inference process, for example, in data acquisition and model construction \cite{zellner1971introduction, sacks1989design, jaynes1991straight, der2009aleatory, arendt2012quantification, brynjarsdottir2014learning}, or in the discretization and numerical computations of the inverse and forward problems \cite{babuvska1978posteriori, sacks1989design, ainsworth1992procedure, oden1993error, oden2000estimation, babuvska2001finite, kennedy2001bayesian, oden2002estimation, prudhomme2003computable, brenner2007mathematical, babuska2010finite, oden2011control, oden2012introduction}. \\

Uncertainty in data acquisition and model construction has been traditionally divided into two categories of {\it epistemic} and {\it aleatoric (aleatory)} \cite{pate1996uncertainties, faber2005treatment, der2009aleatory, beck2010bayesian, haukaas2011model, national2012assessing, beven2016facets}. Aleatoric uncertainties are presumed to stem from inherent {\it unpredictable} variabilities and randomness in observational data and are therefore thought to be irreducible. For example, the experiment of throwing a die could be considered as an experiment with aleatoric uncertainty in its outcome (but note that this statement is incorrect within the Bayesian framework. See \S\ref{sec:introInadAlea} for clarification). This form of uncertainty is sometimes also called {\it structural variability} or {\it risk} \cite{pate1996uncertainties, der2009aleatory} in engineering literature. To the contrary, epistemic uncertainties represent any lack of knowledge about data/experiment that can be potentially acquired in future, for example, the measurement errors in an experiment. \\

There is a long history of confusion and disagreement in the scientific literature about the exact definition and extent, or even the existence of these two forms of uncertainties \cite{jaynes1989clearing, jaynes2003probability}. The origin of this century-long debate can be uncovered in the correspondence that is generally assumed between the two classes of aleatoric vs. epistemic uncertainties and the two prominent interpretations of probability: frequentist vs. Bayesian, respectively \cite{pate1996uncertainties, yager2008classic, chen2017uncertainty}. Thus, from a historical perspective, aleatoric uncertainty is solely defined within the framework of frequentist statistics \cite{chen2017uncertainty}. \\

We remark that in a pure Bayesian system of logical probability \cite{jaynes2003probability}, which is the view we adopt in this work, all uncertainty is epistemic \cite{chen2017uncertainty}: randomness is strictly a means to express a lack of knowledge. By contrast, what is often meant by aleatoric uncertainty in contemporary scientific literature appears to conform well to the concept of {\it model inadequacy}, which is further described below in \S\ref{sec:introInadAlea}. \\

Of course, on scales relevant to Quantum Mechanics \cite{dirac1981principles}, one may argue that Heisenberg's Uncertainty Principle \cite{heisenberg1985anschaulichen} dictates an inherent uncertainty in Nature, setting a hard limit on the extent of human knowledge. This strict epistemological limitation on human knowledge would, therefore, resemble aleatoric uncertainty in Natural phenomena at the ontological level. However, counter-arguments have been put forth by prominent physicists in the $20$th century against this Copenhagen interpretation of Quantum Mechanics \cite{einstein1935can, bohm1952suggested, jaynes1989clearing}. Regardless of the validity of Copenhagen interpretation and the Heisenberg Uncertainty Principle, the quantum mechanical limitations imposed on human knowledge can be considered irrelevant to virtually all practical modeling problems beyond the subatomic scales of Quantum Mechanics (See \cite{rosenthal2006struck} for some illustrative discussions on this topic). \\

\subsection{Model Inadequacy vs. Aleatoric Uncertainty}
\label{sec:introInadAlea}

In an ideal and deterministic world, where there is neither model imperfection nor uncertainty in computation or data, one would naturally expect the physical model to perfectly describe observational data. This is, however, never the case in virtually all real-world inference problems. In reality, all models are imperfect or wrong (echoing the famous statement of E. Box that ``all models are wrong but some can be useful'' \cite{box1976science}) and none can provide a full description of data. This model imperfection is widely known in the literature as {\it model discrepancy} or {\it model inadequacy} \cite{oden2004predictive, national2012assessing}, the possible remedies of which have been already extensively studied \cite{kennedy2001bayesian, arendt2012quantification, brynjarsdottir2014learning, ling2014selection, morrison2016representing}. \\

Model inadequacy is often confused with aleatoric uncertainty since both can have identical effects on scientific inference. From a Bayesian perspective, one can argue that any type of intrinsic unexplained variability observed in natural phenomena is a result of our limited knowledge/data or a consequence of an imperfect physical model for the observed phenomenon. \\

Consider as an example, the experiment of throwing an unbiased die repeatedly under `similar conditions'. What do we really mean here by {\it similar conditions}? Indeed, if the experimenter had {\it complete knowledge} of the conditions under which the die was thrown, there would be no intrinsic unexplained randomness in the experiment's output; that is, one would be able to predict exactly the outcome of each die-throwing experiment. \\

Therefore, our ignorance of the many details of input data to a sequence of experiments and the conditions under which the experiments were performed, manifest themselves in the form of an inherent variability in the experimental output. In other words, no two experiments can be truly considered as identical replicates of each other, because our knowledge of the experimental setup as well as the input data to the experiments is almost always incomplete. This lack of a complete detailed knowledge of the problem under study often leads to the development of mathematical models that are inadequate in describing the underlying physics of the problem correctly. \\

In sum, model inadequacy appears to be frequently confused with aleatoric uncertainty in contemporary scientific literature. From a Bayesian viewpoint, aleatoric uncertainty does not exist. As a result, aleatoric uncertainty is sometimes implicitly redefined as the class of uncertainties for which there is no foreseeable possibility of elimination or reduction {\it at the time of inference} \cite{der2009aleatory}, although it may be reduced with the arrival of new models, experimental designs, or more detailed experimental data in future (see \cite{shafer2008non} for an elegant historical/philosophical review). \\

\subsection{The Goal of This Paper}
\label{sec:introGoal}

Regardless of the terminology used for uncertainty classification, the truth is always convolved with uncertainties that are either due to measurement errors or incomplete (insufficiently-detailed) data. This lack of knowledge, in turn, leads to the development of imperfect physical models whose predictions are inadequate for a complete description of the observed data. As a result, new physically-inspired stochastic and/or deterministic models are needed to further describe the inadequacy of the physical models. \\

Description of a general framework for incorporating different sources of uncertainty, in particular, the measurement error and model inadequacy in the process of scientific inference, seems to be lacking in the current scientific and engineering literature. Most of the few resources available on this topic focus on special cases where the errors involved in the problem take simple Gaussian forms and are assumed to be additive \cite{higdon2004combining, oden2004predictive, clark2006hierarchical, qian2008bayesian, higdon2008computer, arendt2012quantification, brynjarsdottir2014learning}. \\

In the following sections, we consider the effects of model inadequacy (or as sometimes referred to it by `aleatoric uncertainty' in the literature), as well as the effects of noise and measurement error in experimental data, on parameter estimation and predictive inference. Although we have argued, and assume throughout the rest of this paper, that all uncertainties are epistemic (i.e., due to lack of knowledge), we recognize and show in the following sections that model inadequacy and measurement errors require fundamentally different treatments in the process of scientific inference. This is in agreement with the general consensus in the literature \cite{ferson1996different}. A complete description of all the variables used in this manuscript is given in Table \ref{tab:symbols}.

\section{Modeling the Truth}
\label{sec:phys}

\subsection{The Relationship Between Observational Data and the Truth}
\label{sec:physDataVersusTruth}

    Consider a set of $\ndo$ observations,
    \begin{equation}
        \label{eq:dataset}
        \dataset=\{\data_1,\ldots,\data_\ndo\} ~,
    \end{equation}

    \noindent collected about a set of natural phenomena. This observational dataset is a result of the convolution of the unknown reality with various forms of uncertainties in the measurement process. Let $\truth_i$ represent the reality corresponding to the $i$th observation, $\data_i$, in dataset $\dataset$. Let,
    \begin{equation}
        \label{eq:truthset}
        \truthset=\{\truth_1,\ldots,\truth_\ndo\} ~,
    \end{equation}

    \noindent represent the collection of $\ndo$ realities, corresponding to the observational dataset, $\dataset$, as illustrated in Figure \ref{fig:reality2data}. \\

    Each observation, $\data_i\in\dataset$, and the corresponding reality to it, $\truth_i\in\truthset$, is characterized by $\ndv$ variables (\ie~observable quantities), describing different characteristics of the events. Thus, $\truth_i$ is a vector of $\ndv$ elements representing a single event in the $\ndv$-dimensional observational sampling space $\sspace\subset\mathbb{R}^\ndv$. \\

    For the moment (and throughout \S\ref{sec:phys} and \ref{sec:physInad}), suppose we live in an ideal world where natural phenomena are observed and measured exactly and accurately, without any possible bias, inadequacy, or contamination with noise. Therefore, the observational dataset, $\dataset$, represents the reality, $\truthset$, exactly and deterministically. Hence, we will only consider modeling of the truth in this section, dealing only with $\truthset$ and its members, $\truth_i$. \\

    \begin{figure}[t!]
        \begin{minipage}[c]{0.39\textwidth}
            \centering
            {\small
            \begin{tabular}{c"ccccc}
                $\truthset$   & $\truthind$  & $\truthdep$  \\ \thickhline
                $\truth_1$    & $\tvarind[1]{1},\cdots,\tvarind[1]{j}$    & $\tvardep[1]{j+1},\cdots,\tvardep[1]{\ndv}$    \\ \hline
                $\truth_2$    & $\tvarind[2]{1},\cdots,\tvarind[2]{j}$    & $\tvardep[2]{j+1},\cdots,\tvardep[2]{\ndv}$    \\ \hline
                $\vdots$      & $\vdots$ & $\vdots$    \\ \hline
                $\truth_\ndo$    & $\tvarind[\ndo]{1},\cdots,\tvarind[\ndo]{j}$    & $\tvardep[\ndo]{j+1},\cdots,\tvardep[\ndo]{\ndv}$    \\ \hline
            \end{tabular}
            \captionof{table}{{\bf The Truth Set.} The set of all physical events under study.} \label{tab:truthset}
            }
        \end{minipage}
        \begin{minipage}[b]{0.20\textwidth}
            \centering
            \begin{equation*}
            {\Large \textcolor{red}{\mathbf{\xrightarrow[\text{Measurement Error}]{\text{Convolution with}}}}}
            \end{equation*}
        \end{minipage}
        \begin{minipage}[c]{0.39\textwidth}
            \centering
            {\small
            \begin{tabular}{c"ccccc}
                $\dataset$   & $\dataind$  & $\datadep$  \\ \thickhline
                $\data_1$    & $\dvarind[1]{1},\cdots,\dvarind[1]{j}$    & $\dvardep[1]{j+1},\cdots,\dvardep[1]{\ndv}$    \\ \hline
                $\data_2$    & $\dvarind[2]{1},\cdots,\dvarind[2]{j}$    & $\dvardep[2]{j+1},\cdots,\dvardep[2]{\ndv}$    \\ \hline
                $\vdots$      & $\vdots$ & $\vdots$    \\ \hline
                $\data_\ndo$    & $\dvarind[\ndo]{1},\cdots,\dvarind[\ndo]{j}$    & $\dvardep[\ndo]{j+1},\cdots,\dvardep[\ndo]{\ndv}$    \\ \hline
            \end{tabular}
            \captionof{table}{{\bf The Data Set.} A result of the convolution of the truth set with measurement error.} \label{tab:dataset}
            }
        \end{minipage}
        \captionof{figure}{\normalfont{\bf The Experimental Measurement Process.} A schematic representation of the process of collecting observational dataset, $\dataset$, which is a result of the convolution of the unknown truth, $\truthset$, with various sources of uncertainty and measurement error during the experimental data collection process. In the absence of noise and measurement error, the set of observational data, $\dataset$, in Table \ref{tab:dataset} would be identical to the truth set, $\truthset$, in Table \ref{tab:truthset}. Each physical event, $\truth_i$, and the corresponding data observation, $\data_i$, is composed of a set of $\ndv$ event attributes, a subset of which ($\truthdep[i]$ and $\datadep[i]$) are hypothesized/known to depend on the remaining $j$ independent characteristics of the event, ($\truthind[i]$ and $\dataind[i]$). For example, in many problems, the temporal and spatial coordinates are the independent attributes of data and other physical quantities are modeled as a function of these independent attributes.} \label{fig:reality2data}
    \end{figure}

    Although frequently independent of each other, the attributes of each event, $\truth_i$, or its occurrence could also depend on any subset, $\truthsubset[i]\subset\truthset$, of the other events. In general, there may also exist interdependencies between the attributes of each event (\ie~the elements of each vector $\truth_i$). A well-known generic problem of this type in engineering and natural sciences is regression, where the experimenter/observer has control over some characteristics of the events. These characteristics serve as input to the experiment and result in some experimental output that represent the response characteristics of the corresponding events. \\

    Therefore, the attributes of a physical event can be often divided into a set of independent characteristics, $\truth_{ind}$, on which the rest of the event's attributes (\ie~the dependent/response variables, $\truth_{dep}$) depend,
    \begin{equation}
        \label{eq:controlResponseVariables}
        \truth_i=\big\{ ~ \truthind[i] ~ , ~ \truthdep[i] \big( \truthind[i] \big) ~ \big\} ~.
    \end{equation}

    Such modeling scenarios are abundant in science and engineering \cite{oden2004predictive}; for example, a set of fatigue experiments designed to measure maximum tolerable stress (the dependent variable) in a material as a function of strain (the independent variable) \cite{babuvska2016bayesian}, the measurement of the growth of a malignant tumor as a function of time in a murine subject or patient \cite{oden2013selection,hormuth2017mechanically}, the evolution of protein amino acid sequence as function of its structural characteristics \cite{shahmoradi2014predicting, shahmoradi2015dissectingb, shahmoradi2016dissecting, jackson2016intermediate}, or modeling the energetics and occurrence rates of astrophysical phenomena as a function of their distances from the earth \cite{shahmoradi2009real, shahmoradi2010hardness, shahmoradi2011possible, shahmoradi2013gamma, shahmoradi2013multivariate, shahmoradi2015short}.

\subsection{The Forward Problem}
\label{sec:physForwardProblem}

\begin{figure}[t!]
        \centering
        \begin{subfigure}[t]{0.45\textwidth}
            \centering
            \includegraphics[width=\textwidth]{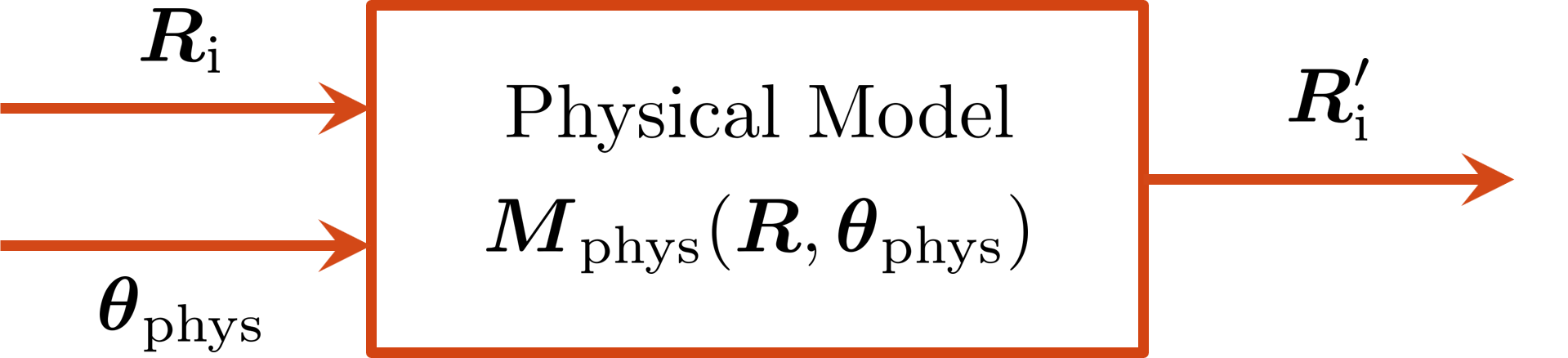}
            \caption{Misspecified model.} \label{fig:physicalModelMisSpecified}
        \end{subfigure}
        \hfill
        \begin{subfigure}[t]{0.45\textwidth}
            \centering
            \includegraphics[width=\textwidth]{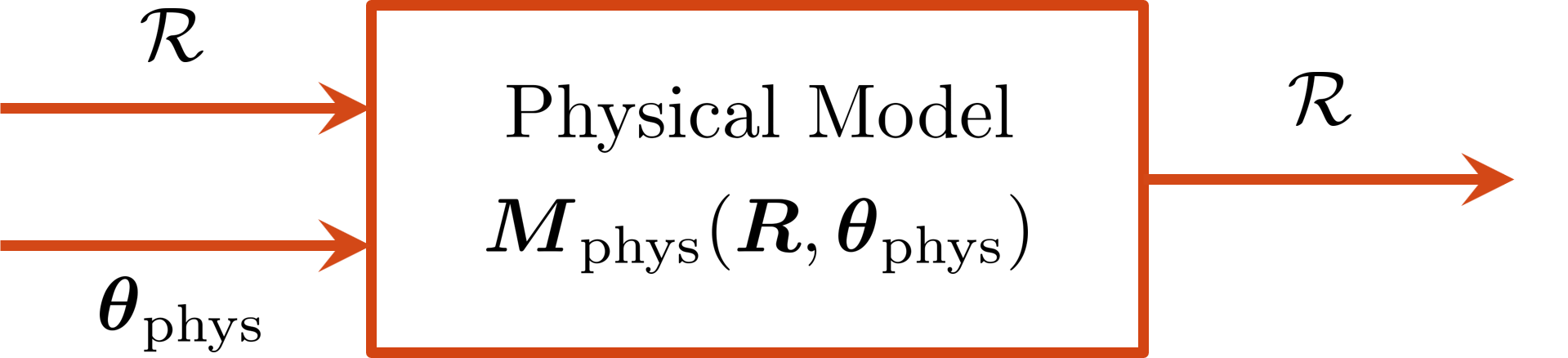}
            \caption{Correctly-specified model.} \label{fig:physicalModelWellSpecified}
        \end{subfigure}
        \caption
        {
            {\bf Schematic diagrams of the physical model with its input and output arguments.} The physical model takes as input a set of values for the model parameters, $\pparam$, as well as the set of attributes, $\truth$, of an event or a collection of events. It then outputs a vector, $\truth'$, of the same length/size as $\truth$, whose elements may or may not be identical to the corresponding elements in $\truth$, depending on $\pparam$ and the validity of the physical model.
            {\bf(a)} A {\bf misspecified physical model} whose output for {\it at least} one event, $\truth_i$, in the event set, $\truthset$, does not match the input vector, $\truth_i$, for any possible values of $\pparam$.
            {\bf(b)} A {\bf correctly-specified physical model} for which there is at least one set of input parameter values, $\pparamopt$, that result in an output from the physical model which is identical to the input, $\truth_i$, for all $\truth_i\in\truthset$ \cite{oden2004predictive}.
        }
        \label{fig:physicalModel}
    \end{figure}

    Now, suppose we formulate a hypothesis with regards to the set of $\ndo$ events in $\truthset$. This hypothesis can be cast in the form of a mathematical model, $\pmodel$, with the subscript $\phys$ emphasizing the type of the model (\ie~a physics-based model). This physical model can be thought of as a collection of mathematical operators (\eg~algebraic, differential, integral, \ldots) that takes as input, a set of $\npp$ physical parameters represented by the vector $\pparam\in\ppspace\subset\mathbb{R}^\npp$. It then acts on some or all $\ndv$ attributes of an independent event or a set of dependent events, $\truth\in\sspace$, that is input to the model and generates an output response, $\truth'$, of the same length/size as $\truth$, which may or may not be identical to the input $\truth$ (Figure \ref{fig:physicalModel}),
    \begin{equation}
        \label{eq:forwardProblem}
         \pmodel \big( \truth, \pparam, \pscen \big) = \truth' ~~~ \forall ~ \truth \subset \sspace ~,
    \end{equation}

    Given the set of events, $\truthset$, as in \eqref{eq:truthset}, the forward problem can be written as,
    \begin{equation}
        \label{eq:forwardProblemTruthset}
         \pmodel \big( \truthset, \pparam, \pscen \big) = \truthset' ~~~ \forall ~ \truthset \subset \sspace ~.
    \end{equation}

    \noindent where,
    \begin{equation}
        \label{eq:physModelOutput}
        \truthset' = \{ \truth_1' , \ldots , \truth_\ndo' \} ~,
    \end{equation}
    \noindent is the output of the physical model, $\pmodel$, corresponding to the event set, $\truthset$. When the events in $\truthset$ are independent of each other, the forward problem takes the more simplified form,
    \begin{equation}
        \label{eq:forwardProblemTruth}
         \pmodel \big( \truth_i, \pparam, \pscen \big) = \truth_i' ~~~ \forall ~ \truth_i \in \truthset \subset \sspace ~.
    \end{equation}

    In general, the observational information about each event, $\truth_i$, might be collected under a specific set of conditions collectively named here as the {\it physical scenario}, $\pscen$. The scenario, $\pscen$, describes the set of all features of the scientific problem at hand, that can be exactly specified, that is also independent of the data, $\truthset$, the physical model, $\pmodel$, and its parameters, $\pparam$. These features typically describe and depend on the experimental/observational setup for data collection \cite{oden2004predictive}. \\

    For example, $\pscen$ could describe the limitations of a detector that was used for data collection \cite{shahmoradi2009real, shahmoradi2010hardness, shahmoradi2011cosmological, shahmoradi2011possible, shahmoradi2013gamma, shahmoradi2013multivariate, shahmoradi2014classification, shahmoradi2015short, shahmoradi2015short}. Thus, although $\pscen$ is independent of $\truthset$, the observation of individual events in the dataset, $\dataset$, and hence the set $\truthset$, is not independent of $\pscen$ in general.

\subsection{The Inverse Problem}
\label{sec:physInverseProblem}

    If the proposed model is capable of describing the phenomenon of interest perfectly without any bias or inadequacy, then by definition, there should exist at least one set of feasible parameter values $\pparam=\pparamopt\in\ppspace$ for which,
    \begin{equation}
        \label{eq:physModel}
        \truthset - \pmodel \big( \truthset, \pparamopt, \pscen \big) = \bs 0 ~,
    \end{equation}

    \noindent where $\bs 0$ is the null matrix of size $\ndo\times\ndv$. In the case of independent events,
    \begin{equation}
        \label{eq:physModelIID}
        \truth_i - \pmodel \big( \truth_i, \pparamopt, \pscen \big) = \bs 0 ~~~ \forall ~ \truth_i \in \truthset ~,
    \end{equation}
    
    \noindent where $\bs 0$ is now the null vector of length $\ndv$. Thus, given a perfect model and an ideal dataset with no measurement error, the problem of inference is reduced to solving \eqref{eq:physModel} or a system of $\ndo$ equations of the form \eqref{eq:physModelIID} to obtain the feasible values, $\pparamopt$, of the set of parameters of the physical model satisfying \eqref{eq:physModel} or \eqref{eq:physModelIID}. \\

    In mathematical modeling, this process is widely known as {\it inverse problem} or {\it model calibration}, as illustrated in Figure \ref{fig:scientificMethod}. When there is a $\pparamopt$ for which the physical model, $\pmodel$, perfectly describes the truth (by satisfying \eqref{eq:physModel} or \eqref{eq:physModelIID}), then $\pmodel$ is said to be a {\it correctly-specified} or {\it well-specified} model \cite{oden2004predictive} as illustrated in Figure \ref{fig:physicalModelWellSpecified}. In some problems, there might exist multiple or even uncountably infinite number of $\pparamopt$ that satisfy \eqref{eq:physModel}. In such cases, the system of equations is said to be degenerate or ill-posed in the sense of Hadamard \cite{hadamard1902problemes}. 
\section{Modeling the Truth in the Presence of Model Inadequacy}
\label{sec:physInad}

One can envisage many experimental setups in which the same input to the experiment, results in a variety of possible outcomes. For example, the experiment of throwing an unbiased die under similar conditions yields six inherently different possible outcomes. A more relevant example to the field of engineering is a set of stress-strain data obtained from a heterogeneous material. In such data, a single strain could yield several different stress values in repeated identical experiments, depending on the level of heterogeneity and imperfection of the material being tested. \\

Such apparently unexplained variabilities and heterogeneities in observational data with respect to the predictions of the physical model, fall into the category of {\it model inadequacy}, or {\it model discrepancy}, sometimes also referred to it as {\it structural uncertainty} or {\it aleatoric uncertainty}. This is a class of uncertainty that presumably cannot be reduced with further collection of data or more accurate measurements of {\it the same characteristic features} already describing each event, $\truth_i\in\truthset$. \\

Regardless of the origins of model inadequacy -- whether it be a wrong physical model or insufficiently-detailed data -- such discrepancies between model predictions and data almost always exist, and their behavior has to be described by yet another set of models of stochastic origin. \\

\subsection{The Construction of Inadequacy Model}
\label{sec:physInadInadModel}

    When there is unexplained variability in the events with respect to the predictions of the physical model, the model is said to be {\it inadequate} or {\it misspecified} \cite{oden2004predictive}, and the output of the physical model for some or all events, $\truth_i\in\truthset$, does not match the available data. Hence, the equalities in \eqref{eq:physModel} and \eqref{eq:physModelIID} do not hold. Rather we have in general,
    \begin{eqnarray}
        \label{eq:physInadResidual}
        \truthset - \pmodel \big( \truthset, \pparam , \pscen \big) = \randset ~,
    \end{eqnarray}

    \noindent where $\randset$ is set of $\ndo$ vectors, $\rand_i$, each of which corresponds to one event, $\truth_i\in\truthset$. In the case of independent events, \eqref{eq:physInadResidual} takes the more simplified form,
    \begin{eqnarray}
        \label{eq:physInadResidualIID}
        \truth_i - \pmodel \big( \truth_i, \pparam , \pscen \big) = \rand_i ~~~ \forall ~ \truth_i \in \truthset ~.
    \end{eqnarray}

\begin{figure}[t!]
        \begin{center}
            \includegraphics[width=\textwidth]{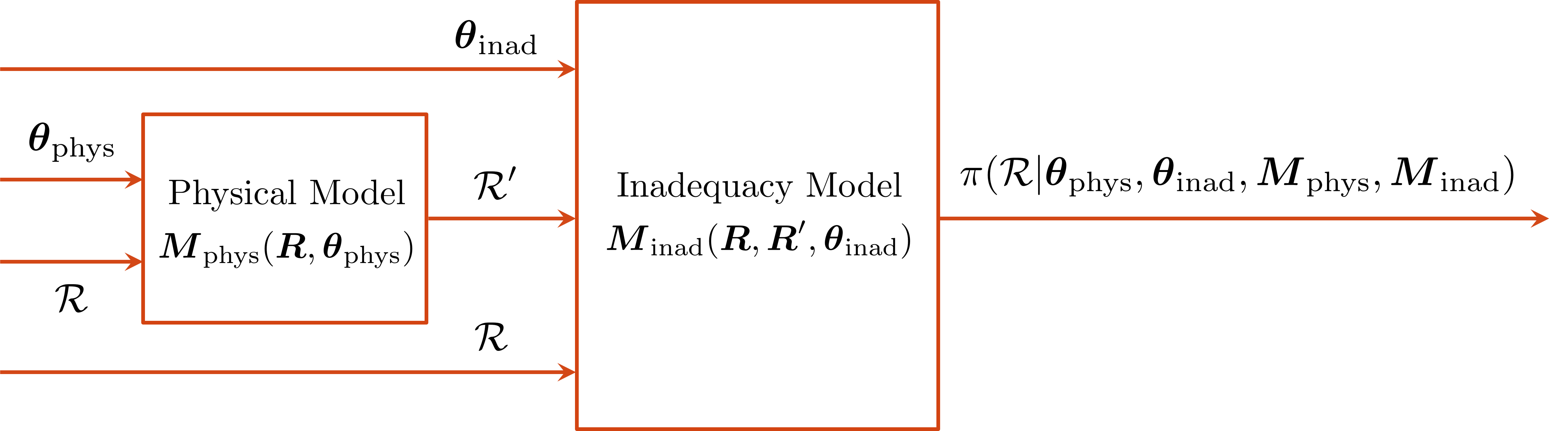}
        \end{center}
        \caption
        {
            A schematic illustration of the relationship between the physical model, $\pmodel$, and the inadequacy model, $\amodel$: the statistical model that describes the inadequacy of the physical model with respect to the available data. On input, $\amodel$ takes a set of values for its parameters, $\aparam$, the output of the physical model, $\truthset'$, and the truth about the collection of events, $\truthset$. On output, $\amodel$ gives the probability of the truth being $\truthset$, given the physical model's output, $\truthset'$, and the specific values of the parameters of the physical and inadequacy models: $\{\pparam,\aparam\}$.
        }
        \label{fig:physInadLike}
    \end{figure}

    Each $\rand_i$ is a vector of length $\ndv$, some elements of which, corresponding to the response variables, $\truthdep[i]\subset\truth_i$, are not anymore deterministic but random stochastic variables. In other words, for the same control variables $\truthind\subset\truth=\{\truthdep,\truthdep\}$, there can be a finite or infinite number of possible values for the response vector $\truthdep$. But the physical model is only capable of fitting some representative deterministic average output response to $\truthdep[i]\subset\truth_i~,~\forall\truth_i\in\truthset$. Therefore, the set,
    \begin{equation}
        \label{eq:inadRandset}
        \randset=\{\rand_{1},\ldots,\rand_{\ndo}\} ~,
    \end{equation}

    \noindent is a collection (\ie~ a $\ndo\times\ndv$ matrix) of deterministic and stochastic variables whose behavior has to be determined by yet another model, $\amodel$, of statistical and physical origin (as opposed to the deterministic physical origin of $\pmodel$ (Figure \ref{fig:physInadLike}). This stochastic model depends explicitly on the output of the proposed physical model, $\pmodel$,
    \begin{eqnarray}
        \label{eq:physInadModel}
        \randset \sim \amodel \big( \truthset, \aparam, \pmodel ( \truthset, \pparam , \pscen ) \big) ~,
    \end{eqnarray}

    \noindent where $\aparam$ represents the set of $\npa$ parameters of $\amodel$. Here, the subscript $\alea$ stands for {\it inadequacy}. In the case of independent and identically distributed (\iid) events, the above equation can be written as,
    \begin{eqnarray}
        \label{eq:physInadModelIID}
        \rand_i \sim \amodel \big( \truth_i, \aparam, \pmodel ( \truth_i, \pparam , \pscen ) \big) ~.
    \end{eqnarray}

    The special case where the parameters of $\amodel$ depend explicitly on $\truthind[i]\subset\truth_i$, is known as {\it heteroscedasticity} in the statistical literature \cite{pearson1904mathematical}. \\

\subsection{The Likelihood Function in the Presence of Model Inadequacy}
\label{sec:physInadLikelihood}

    In the presence of model inadequacy, the inference problem cannot be solved deterministically anymore as in \S\ref{sec:physInverseProblem}, rather one has to first derive the likelihood function (i.e., the unnormalized density function), $\like$, of the parameters of the physical and inadequacy models. Let,
    \begin{eqnarray}
        \pamodel &=& \{\pmodel,\amodel\} ~, \\
        \paparam &=& \{\pparam,\aparam\} ~,
    \end{eqnarray}

    \noindent represent respectively, the combined physical and inadequacy models for $\truthset$, and the combined set of parameters of the two models. Then, by definition,
    \begin{eqnarray}
        \pi \big( \truthset \big| \paparam , \pamodel \big) \equiv \pi \big( \randset \big| \paparam, \pamodel \big) ~, \label{eq:physInadModelPDF}
    \end{eqnarray}
    represents the multivariate probability density function (PDF), $\pi(\cdot)$, of obtaining $\truthset$ given $\pamodel$ and the parameter values $\pamodel$, which in the case of \iid~events simplifies to,
    \begin{eqnarray}
        \pi \big( \truth_i \big| \paparam , \pamodel \big) \equiv \pi \big( \rand_i \big| \paparam, \pamodel \big) 
         ~~~ \forall ~ \truth_i \in \truthset
        ~,
        \label{eq:physInadModelPDFIID}  ~.
    \end{eqnarray}

    Then, the likelihood of $\paparam$ being the correct set of values for the parameters of $\pamodel$, in the light of available data, $\truthset$, becomes,
    \begin{eqnarray}
        \like \big( \paparam ; \truthset \big)
        &\equiv&     ~ \like \big( \paparam ; \randset \big) \label{eq:physInadLike} \\
        &\equiv&     ~ \pi \big( \truthset \big| \paparam , \pamodel \big) \label{eq:physInadLikeJoint} \\
        &\ceq{\iid}& ~ \prod_{i=1}^{\ndo} \pi \big( \truth_i \big| \aparam , \amodel \big( \truth_i, \aparam, \pmodel(\truth_i,\pparam,\pscen) \big) \big) \label{eq:physInadLikeIID} ~,
    \end{eqnarray}

    \noindent where again, the last equality, \eqref{eq:physInadLikeIID}, holds only on the special occasion where the observed events are independent of each other and are equally likely to occur, that is, independent and identically distributed (\iid). Although the \iid~property represents a special case, in practice it holds for a wide range of scientific inference and modeling problems. \\

    The problem of inference is now reduced to finding a $\paparam=\{\pparam,\aparam\}$ such that the joint probability of obtaining all events in $\truthset$ together, combined with any prior knowledge about the parameters of the models, $\pamodel=\{\pmodel,\amodel\}$, is maximized. From the Bayes rule, it follows,
    \begin{eqnarray}
        \label{eq:physInadBayesRand}
        \pdfbig{\paparam \big| \randset, \pamodel, \prior_\paparam} = \frac
        {
            \pdfbig{\randset \big| \paparam , \pamodel} \pdfbig{\paparam \big| \pamodel,\prior_\paparam}
        }
        {
            \pdfbig{\randset \big| \pamodel, \prior_\paparam}
        } ~,
    \end{eqnarray}

    \noindent or equivalently,
    \begin{eqnarray}
        \label{eq:physInadBayes}
        \pdfbig{\paparam \big| \truthset, \pamodel, \prior_\paparam} = \frac
        {
            \pdfbig{\truthset \big| \paparam , \pamodel} \pdfbig{\paparam \big| \pamodel,\prior_\paparam}
        }
        {
            \pdfbig{\truthset \big| \pamodel, \prior_\paparam}
        } ~,
    \end{eqnarray}

    \noindent where $\prior_\paparam$ represents any prior knowledge about all unknown parameters, $\paparam$, of the physical and inadequacy models together. The subject of constructing the prior PDF, $\pdfbig{\paparam|\pamodel,\prior_\paparam}$, from the available knowledge in an inference problem is as old as the Bayesian probability theory itself. Over the past century, several methods such as {\it Jeffreys' principle of invariance under reparametrization} \cite{jeffreys1998theory} or {\it Jaynes' principle of maximum entropy} \cite{jaynes1957informationI, jaynes1957informationII, jaynes1973well, jaynes2003probability, jaynes2003probability, gregory2005bayesian} have been developed to construct objective priors for Bayesian inference.

\subsection{Popular Choices of Inadequacy Model}
\label{sec:physInadModelChoices}

    The Multivariate Normal (MVN) distribution is undoubtedly the most popular and widely-used choice of probability density function for inadequacy models in scientific inference, although it has appeared under different names, notably the Least-Squares method first introduced by Adrien Legendre in 1805 \cite{legendre1805nouvelles}. Another popular choice of inadequacy model is the Laplace distribution, more commonly known as the Least Absolute Deviation method first introduced by Pierre Laplace in 1774 \cite{laplace2012pierre, keynes1911principal}. \\

    Despite their popularity, neither MVN nor Laplace are the most appropriate choices of inadequacy model, $\amodel$, for every scientific inference problem. Since $\amodel$ represents the inadequacy of the physical model, in general, it has to be also inferred from the characteristics of data and the physical phenomena being investigated. 
\section{Modeling the Truth, Confounded with Noise (Measurement Error)}
\label{sec:physNois}

Independently of the model inadequacy, experimental observations are always contaminated with measurement error (or equivalently as we use hereafter, noise). Such sources of uncertainty in data are sometimes called {\it epistemic uncertainty}, although the scope of epistemic uncertainty goes beyond measurement error, as we discussed in \S\ref{sec:intro}. \\

Unlike model inadequacy, uncertainty due to measurement error is a result of the fundamental limitations of the measurement process and instruments. It is therefore assumed to be reducible by gathering higher quality information about the phenomenon of interest, for example, by making more accurate measurements with more accurate devices. \\

The inevitable existence of noise in every real-world experiment implies that the truth, $\truth$, about an event will never be known to the observer/experimenter unless the uncertainty in data is exactly and {\it deterministically} modeled and removed. This is, however, impossible since the effects of noise on reality are virtually never known deterministically. Under the most optimistic scenarios, an experimenter may only be able to make an educated guess on the general average stochastic effects of the noise on the experimental measurements. \\

In other words, what an observer/experimenter can perceive about the truth, $\truth$, of an event is only the {\it stochastic} output, $\data$, of a complex measurement process whose input is $\truth$. Therefore, a primary and major task in designing and performing an experiment is to alleviate to the extent possible, the effects of this stochastic noise component on inferences made about the truth. \\

Under a well-determined experimental setup and measurement process, one may be able to provide a stochastic model for the random effects of noise on the truth. Let,
\begin{eqnarray}
    \label{eq:noisModel}
    \data_i \sim \emodel[i](\truth_i, \eparam[i] \big) ~,
\end{eqnarray}

\noindent represent the $i$th observational data in the dataset $\dataset=\{\data_1,\ldots,\data_\ndo\}$ which consists of $\ndo$ observations. Here, $\emodel[i]$ represents the stochastic noise model for the $i$th event, $\truth_i$. Each observation, $\data_i\in\dataset$, results from a complex convolution of the truth, $\truth_i\in\truthset$, with various types of noise in the measurement process, all of which we hope to approximate and model by $\emodel[i]$. \\

The noise model, $\emodel[i]$, takes as input, the set of $\npe[i]$ parameters represented by the vector $\eparam[i]$. Here the subscript $i$, wherever it appears in \eqref{eq:noisModel} and throughout this section, is used to indicate that the corresponding object specifically belongs to or is defined in relation with the $i$th event, $\truth_i$. The subscript $\paraChar\episChar$ in $\npe[i]$ stands for the {\it p}arameters of the {\it n}oise model. \\

A typical example of such inference problem, involving distinct noise models corresponding to each $\truth_i\in\truthset$, can be found in the field of Astronomy and Cosmology where observational data is frequently gathered by multiple instruments of different measurement accuracies \cite{shahmoradi2010hardness, shahmoradi2011cosmological, shahmoradi2013multivariate, shahmoradi2015short}. \\

Since $\data_i$ is a random variable, there is {\it no} one-to-one mapping between $\data_i$ and the truth $\truth_i$. Given the noise model, $\emodel[i]$, and its parameters, $\eparam[i]$, the corresponding PDF of $\data_i$ is,
\begin{eqnarray}
    \label{eq:noisModelPDF}
    \text{PDF} \big(\data_i\big) \ceq{def} \pi \big( \data_i \big| \truth_i, \eparam[i], \emodel[i] \big)
\end{eqnarray}


There are multiple reasons that render \eqref{eq:noisModelPDF} useless by itself, even when the exact mathematical form of $\emodel[i]$ as well as its parameters, $\eparam[i]$, are known:
\begin{itemize}
    \item Firstly, this formulation requires us to know the truth $\truth_i$ as an input to $\emodel[i]$. In practice however, we never know the truth.
    \item Secondly, the observer/experimenter can never obtain multiple realizations of $\data_i$ for the exact same truth $\truth_i$. What an experimenter gathers about an event, $\truth_i$, is a single observation of it, $\data_i$, which is a result of the convolution of $\truth_i$ with noise. From a Bayesian perspective, such experiments can never be repeated under the exact same conditions to obtain multiple observations, $\data_i$, for the same $\truth_i$.
    \item Thirdly and most importantly, the input quantities to the physical model, $\pmodel$, in \eqref{eq:physModel} and \eqref{eq:physModelIID} are $\truthset$ and $\truth_i$ respectively, not $\dataset$ and $\data_i$.
\end{itemize}

A better formulation of the problem can be obtained by asking an appropriate question in relation to noise in data: Given a single observation, $\data_i$, obtained for an unknown event, $\truth_i$, what is the probability that the underlying truth about this event is $\ptruth_i$? Here the superscript $\possible$ is to emphasize that $\ptruth_i$ may not necessarily correspond to the truth $\truth_i$. \\

This question has a straightforward answer using the Bayesian inversion method applied to \eqref{eq:noisModelPDF}, such that the probability of $\ptruth_i$ being the truth about the $i$th event could be written as,
\begin{eqnarray}
    \label{eq:truthPosterior}
    \pi \big( \ptruth_i \big| \data_i, \eparam[i], \emodel[i], \prior_{\truth_i} \big) = \frac
    { \pdfbig{ \data_i \big| \ptruth_i, \eparam[i], \emodel[i] } ~ \pdfbig{ \ptruth_i \big| \prior_{\truth_i} } }
    { \pdfbig{ \data_i \big| \eparam[i] , \emodel[i] , \prior_{\truth_i} } } ~,
\end{eqnarray}

\noindent where the left-hand-side of the equation is the posterior probability density function of the truth, the first term in the numerator is the likelihood function of $\ptruth_i$, which is equivalent to the probability density function of $\data_i$ in \eqref{eq:noisModelPDF}, and $\prior_{\truth_i}$ represents the experimenter's prior knowledge about $\truth_i$. The denominator is simply a normalization factor (the Bayesian evidence) that makes the left-hand-side a properly normalized PDF. It gives the probability of observing $\data_i$ averaged over all possible values, $\ptruth_i$, for the truth, $\truth_i$. \\

In practice, the state of experimenter's prior knowledge, $\prior_{\truth_i}$, about the truth is often {\it complete ignorance}, meaning that all possibilities for the truth are equally probable in the eyes of the experimenter, regardless of the noise model. Therefore, $\pdfbig{\ptruth_i\big|\prior_{\truth_i}}$ is frequently assigned an improper (unbounded) uniform distribution \cite{jaynes1957informationI, jaynes1957informationII, jaynes2003probability}, often without explicitly acknowledging it. There are, however, important exceptions to this general rule, for example, in hierarchical Bayesian inference problems \cite{shahmoradi2013multivariate,shahmoradi2015short} where $\truth_i$ is defined only on a subset of the real numbers, $\mathbb{R}$.

\subsection{Noise Models Are Fundamentally Different From Inadequacy Models}
\label{sec:physNoisNoiseVsInad}

    Unlike the case of model inadequacy studied in \S\ref{sec:physInad} where the parameters, $\aparam$, of the inadequacy model, $\amodel$, were a priori unknown and were to be inferred from data, here it is essential to know the mathematical form of $\emodel[i]$ (\ie~the PDF in \eqref{eq:noisModelPDF}) for each observation, $\data_i$, as well as its parameter values, $\eparam[i]$. In other words, the noise model and its parameters are part of observational data that will have to be fed to the physical and inadequacy models. Otherwise, it would be impossible to construct \eqref{eq:truthPosterior}, and subsequently solve \eqref{eq:physModel}. \\

    Another difference with the case of inadequacy model is the fact that the presence of noise in data does not necessarily invalidate \eqref{eq:physModel}. As long as the physical model, $\pmodel$, is perfectly capable of describing $\truthset$ -- an ideal assumption that we make in this section (\ie~\S\ref{sec:physNois}) -- and the sampling space of $\ptruth_i$ is accurately defined such that,
    \begin{eqnarray}
        \label{eq:noisTruthEncompassingCondition}
        \pdfbig{ \truth_i \big| \data_i, \eparam[i], \emodel[i] } \neq 0 ~~~ \forall~\truth_i\in\truthset ~,
    \end{eqnarray}

    \noindent then, there will be at least one set of physical parameters, $\pparamopt$, for which $\ptruthset=\truthset$ satisfies \eqref{eq:physModel},
    \begin{eqnarray}
        \label{eq:physNoisModel}
        \ptruthset - \ptruthset' = \bs 0 ~,
    \end{eqnarray}

    \noindent where,
    \begin{eqnarray}
    \label{eq:forwardProblem*}
        \ptruthset' &=& \pmodel \big( \ptruthset , \pparam , \pscen \big) ~,
    \end{eqnarray}

    \noindent or in the case of \iid~events,
    \begin{eqnarray}
        \label{eq:physNoisModelIID}
        \ptruth_i - \ptruth_i' = \bs 0 ~~~ \forall ~ \ptruth_i \in \ptruthset ~,
    \end{eqnarray}

    \noindent with,
    \begin{eqnarray}
    \label{eq:forwardProblemIID*}
        \ptruth_i' &=& \pmodel \big( \ptruth_i , \pparam , \pscen \big)  ~~~ \forall ~ \ptruth_i \in \ptruthset ~.
    \end{eqnarray}

    The symbol $\ptruthset$ in \eqref{eq:physNoisModel} represents a set of possible realizations, $\ptruth_i$, of each of the events, $\truth_i\in\truthset$,
    \begin{eqnarray}
        \label{eq:ptruthset}
        \ptruthset = \big\{ \ptruth_1 , \cdots , \ptruth_\ndo \big\} ~.
    \end{eqnarray}

    Therefore, the set $\ptruthset$ represents a possible realization of $\truthset$, which may satisfy \eqref{eq:physNoisModel} for no $\pparam$ values, or a single value, or multiple values, or an infinite number of values of $\pparam$. Let,
    \begin{eqnarray}
        \emodelset&=&\{\emodel[1],\ldots,\emodel[\ndo]\} ~, \label{eq:emodelset} \\
        \eparamset&=&\{\eparam[1],\ldots,\eparam[\ndo]\} ~, \label{eq:eparamset}
    \end{eqnarray}

    \noindent denote respectively, the set of noise models corresponding to each event in $\truthset$, and the set of `a priori known' parameters of each model, $\emodel[i]\in\emodelset$. Then, one can write the probability of $\ptruthset$ being the truth, $\truthset$, via the Bayes rule as,
    \begin{eqnarray}
        \centering
        \pdfbig{\ptruthset \big| \dataset, \eparamset, \emodelset, \prior_\truthset}
        &=& \frac
        { \pdfbig{ \dataset \big| \ptruthset, \eparamset, \emodelset } ~ \pdfbig{ \ptruthset \big| \prior_{\truthset} } }
        { \pdfbig{ \dataset \big| \eparamset, \emodelset, \prior_\truthset } } \label{eq:ptruthsetProbabilityJoint} \\
        &\ceq{\iid}& \prod_{i=1}^\ndo \pdfbig{ \ptruth_i \big| \data_i, \eparam[i], \emodel[i], \prior_{\truth_i} } \label{eq:ptruthsetProbabilityIID} \\
        &\ceq{\iid}& \prod_{i=1}^\ndo \frac
        { \pdfbig{ \data_i \big| \ptruth_i, \eparam[i], \emodel[i] } ~ \pdfbig{ \ptruth_i \big| \prior_{\truth_i} } }
        { \pdfbig{ \data_i \big| \eparam[i], \emodel[i], \prior_{\truth_i} } } ~.  \label{eq:ptruthsetProbabilityIIDExpanded}
    \end{eqnarray}

    \noindent where $\prior_{\truthset}$ represents the experimenter's prior knowledge about the entire set $\truthset$. The latter two equalities hold only under the assumption of independent and identical distribution (\iid) of the experimental measurements, $\data_i\in\dataset$.


\subsection{The Likelihood Function in the Presence of Data Uncertainty}
\label{sec:physNoisLikelihood}

    In most practical inference problems, the noise model, $\emodel[i]$, for a given event, $\truth_i\in\truthset$, has a continuous probability density function for $\ptruth_i$, leading to infinitely many possibilities, $\ptruth_i$, for each event, $\truth_i$. This means that the set $\ptruthset$ is not unique. Therefore, one can construct a superset, $\ptruthSuperset$, consisting of all possible combinatorial realizations of the set of events in $\truthset$,
    \begin{eqnarray}
        \label{eq:ptruthSuperset}
        \ptruthSuperset = \big\{ \ptruthset_1 , \ldots , \ptruthset_j , \ldots \big\} ~,
    \end{eqnarray}

    \noindent where {\it only and only} one realization, $\ptruthset_j\in\ptruthSuperset$, corresponds to the truth set, $\truthset$. Many members of $\ptruthSuperset$ might satisfy \eqref{eq:physNoisModel} for some specific parameter values, $\pparam$, that is not necessarily the same parameter values for which $\truthset$ satisfies \eqref{eq:physNoisModel}. To the contrary, there might exist other members of $\ptruthSuperset$ that do not satisfy \eqref{eq:physNoisModel} for any possible parameter values, $\pparam$. \\

    The superset $\ptruthSuperset$ can be finite, or countably/uncountably infinite, depending on the type of measurement uncertainties involved in the problem. When all of the noise models, $\emodel[i]\in\emodelset$, for the specific problem under study give rise to sets of finite possibilities, $\ptruth_i$, for the corresponding events, $\truth_i\in\truthset$, then the size of $\ptruthSuperset$ would be also finite. Otherwise, if any or all of the noise models are countably/uncountably infinite, then $\ptruthSuperset$ would be also countably/uncountably infinite. \\

\begin{figure}[t!]
        \begin{center}
            \includegraphics[width=\textwidth]{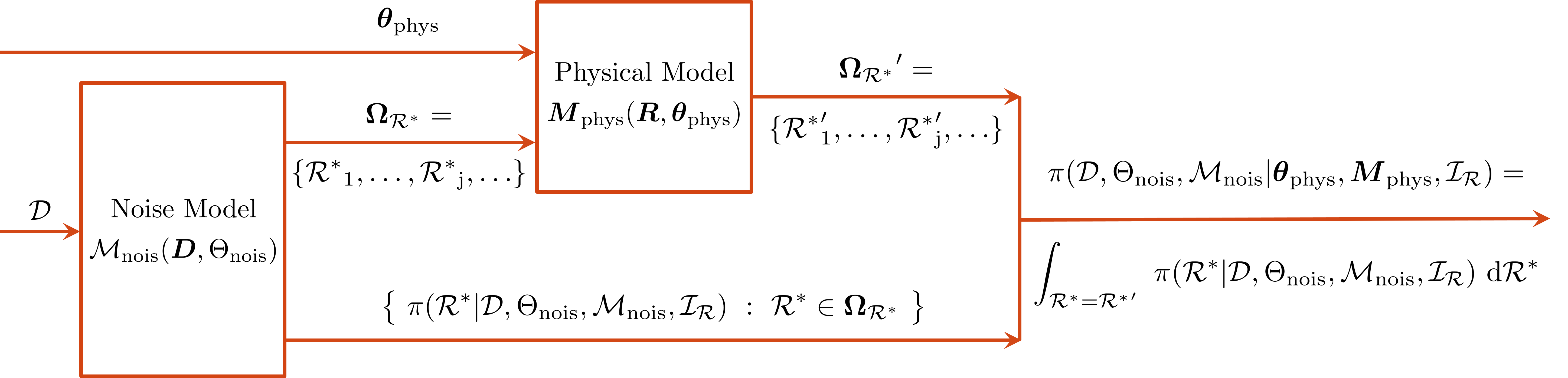}
        \end{center}
        \caption
        {
            A schematic illustration of the derivation of the likelihood function when the physical model, $\pmodel$, has no inadequacy and the observational dataset, $\dataset$, is subject to noise and measurement error. In such scenario, if the set of noise models and their parameters, $\emodelset(\eparamset)$, affecting the data are known, then one has to consider all possibilities, $\{\ptruthset_1,\ldots,\ptruthset_j,\ldots\}$, for the unknown truth, $\truthset$, underlying the observational data, $\dataset$. All such possibilities are then fed to the physical model, which yield the corresponding output set $\{\ptruthset'_1,\ldots,\ptruthset'_j,\ldots\}$. Then, the probability of observing $\dataset$ given a specific set of values for the parameters, $\pparam$, of the physical model, is the sum of the probabilities of all possible realizations, $\ptruthset$,  of the truth, $\truthset$, that satisfy the physical model (\ie~when $\ptruthset=\ptruthset'$).
        }
        \label{fig:physNoisLike}
    \end{figure}

    Now, let $\ptruthSupersetPparam\subset\ptruthSuperset$ represent the set of all possible realizations, $\ptruthset\in\ptruthSuperset$, of $\truthset$ that satisfy \eqref{eq:physNoisModel} for the given $\pparam$. Then, the likelihood (\ie~the unnormalized probability) of $\pparam$ being the true set of parameter values for $\pmodel$ can be written as the sum of the probabilities of all possible realizations, $\ptruthset\subset\ptruthSupersetPparam$,
    \begin{eqnarray}
        \like\big( \pparam \big)
        &\equiv& \pdfbig{ \dataset , \eparamset, \emodelset \big| \pparam , \pmodel, \prior_\truthset } \label{eq:physNoisDataPDF} \\
        &=& \int_{\ptruthSupersetPparam} \pdfbig{ \ptruthset \big| \dataset, \eparamset, \emodelset, \prior_\truthset } ~ \diff\ptruthset \label{eq:physNoisParamLike} \\
        &=& \int_{\ptruthSuperset} \indic\big(\ptruthset\big) ~ \pdfbig{ \ptruthset \big| \dataset, \eparamset, \emodelset, \prior_\truthset } ~ \diff\ptruthset \label{eq:physNoisParamLikeIndic} \\
        &\ceq{\iid}& \prod_{i=1}^\ndo ~ \pdfbig{ \data_i , \eparam[i], \emodel[i] \big| \pparam , \pmodel, \prior_{\truth_i} } \label{eq:physNoisDataPDFIID} \\
        &\ceq{\iid}& \prod_{i=1}^\ndo ~ \int_{\ptruthSuperset} \indic\big(\ptruth_i\big) ~ \pdfbig{ \ptruth_i \big| \data_i, \eparam[i], \emodel[i], \prior_{\truth_i} } ~ \diff\ptruth_i ~, \label{eq:physNoisParamLikeIndicIID}
    \end{eqnarray}

    \noindent where $\indic(\cdot)$ is an indicator function which has the value $1$, only if $\ptruthset$ and $\ptruth_i$ satisfy \eqref{eq:physNoisModel} and \eqref{eq:physNoisModelIID} respectively for the given value of $\pparam$, otherwise it is 0,
    \begin{eqnarray}
        \label{eq:identityFunc}
        \indic\big(\ptruth_i\big)
        &\ceq{def}&
        \begin{cases}
            1 ~~~&~~~ \ptruth_i' = \ptruth_i ~, \\
            0 ~~~&~~~ \text{otherwise} ~,
        \end{cases}
        ~~~ , ~~~ 1 < i < \ndo ~, \\
        \label{eq:identityFunc}
        \indic\big(\ptruthset\big)
        &=& \prod^\ndo_{i=1} ~ \indic\big(\ptruth_i\big) \\
        &\ceq{def}&
        \begin{cases}
            1 ~~~&~~~ \ptruthset' = \ptruthset ~, \\
            0 ~~~&~~~ \text{otherwise} ~.
        \end{cases}
    \end{eqnarray}

    \noindent where $\ptruthset'$ and $\ptruth'_i$ are defined by \eqref{eq:forwardProblem*} and \eqref{eq:forwardProblemIID*}. The equality in \eqref{eq:physNoisParamLikeIndicIID} holds only under the assumption of independent and identical distribution (\iid) of the experimental measurements, $\data_i\in\dataset$. A schematic diagram illustrating the derivation of this likelihood function is given in Figure \ref{fig:physNoisLike}. \\

    Thus, the problem of inference is now to compute and maximize the PDF of $\pparam$, which can be obtained by combining \eqref{eq:physNoisParamLike} with any prior knowledge, $\prior_\pparam$, about the parameters of the model using the Bayes rule,
    \begin{eqnarray}
        \pi \big( \pparam &\big|& \dataset, \eparamset, \emodelset, \pmodel, \prior_\truthset, \prior_\pparam \big) \nonumber \\
        &=& \frac
        {
            \pdfbig{ \dataset , \eparamset, \emodelset \big| \pparam , \pmodel, \prior_\truthset }
            ~ \pdfbig{\pparam \big| \pmodel,\prior_\pparam}
        }
        {
            \pdfbig{ \dataset, \eparamset, \emodelset \big| \pmodel, \prior_\truthset, \prior_\pparam}
        } \label{eq:physNoisBayes1} \\
        &=& \frac
        {
            \bigg[ \bigints_{\ptruthSupersetPparam} \pdfbig{ \ptruthset \big| \dataset, \eparamset, \emodelset, \prior_\truthset } ~ \diff\ptruthset \bigg]
            ~ \pdfbig{\pparam \big| \pmodel,\prior_\pparam}
        }
        {
            \pdfbig{ \dataset, \eparamset, \emodelset \big| \pmodel, \prior_\truthset, \prior_\pparam}
        } ~, \label{eq:physNoisBayes2}
    \end{eqnarray}

    \noindent where the denominator is simply a factor that properly normalizes the posterior distribution of $\pparam$ to a proper probability density function, such that the integral of \eqref{eq:physNoisBayes2} over the entire parameter space, $\pspace_\phys$, is $1$,
    \begin{eqnarray}
        \label{eq:physNoisBayesEvidence}
        && \pdfbig{ \dataset, \eparamset, \emodelset \big| \pmodel, \prior_\truthset, \prior_\pparam} = \nonumber \\
        && \int_{\pspace_\phys}
        \bigg[ \int_{\ptruthSupersetPparam} \pdfbig{ \ptruthset \big| \dataset, \eparamset, \emodelset, \prior_\truthset } ~ \diff\ptruthset \bigg]
        ~ \pdfbig{\pparam \big| \pmodel,\prior_\pparam}
        ~ \diff\pparam
        ~.
    \end{eqnarray}

    It is clear from \eqref{eq:physNoisParamLike}, \eqref{eq:physNoisParamLikeIndic}, \eqref{eq:physNoisParamLikeIndicIID}, and \eqref{eq:physNoisBayes2}, that the posterior PDF of the parameters of the physical model, $\pdfbig{ \pparam \big| \dataset, \eparamset, \emodelset, \pmodel, \prior_\pparam }$, depends only implicitly on the physical model, $\pmodel$, and its parameters, $\pparam$. The only influence of $\pmodel$ and $\pparam$ on the posterior PDF of \eqref{eq:physNoisBayes2} is through the definition of the domain of integration in the likelihood function of $\pparam$ in \eqref{eq:physNoisParamLike}, \eqref{eq:physNoisParamLikeIndic}, and \eqref{eq:physNoisParamLikeIndicIID}. \\
    
    This seemingly-bizarre behavior of the posterior PDF in \eqref{eq:physNoisBayes2} is a natural consequence of the underlying idealistic assumption that we have made in this section, that is, there is no model inadequacy and the physical model could perfectly describe $\truthset$ if we knew it. 
\section{Hierarchical Modeling of the Truth, in the Presence of Model Inadequacy and Data Uncertainty}
\label{sec:physInadNois}

So far, none of the idealized scenarios in \S\ref{sec:phys}, \S\ref{sec:physInad}, and \S\ref{sec:physNois} represent what a practical researcher confronts in modeling natural phenomena. In reality, an experimenter collects a dataset, $\dataset$, each event of which is,
\begin{enumerate}
    \item contaminated with various types of noise (measurement error) and is also,
    \item insufficiently detailed,
\end{enumerate}
\noindent leading to the development of wrong or incomplete physical models that are inadequate in providing a full description of the available dataset, $\dataset$. Hence, the resulting dataset appears to be heterogeneous with respect to the predictions of the physical model at hand.

\subsection{Statement of the Hierarchical Bayesian Inverse Problem}
\label{sec:physInadNoisProblem}

    Given,

    \begin{enumerate}
         \item the dataset $\dataset=\big\{\data_i,\cdots,\data_\ndo\big\}$,
         \item the corresponding set of noise models, $\emodelset$, and their `a priori known' parameters, $\eparamset$, as given by \eqref{eq:emodelset} and \eqref{eq:eparamset}, corresponding to each observation, $\data_i\in\dataset$,
         \item a physical model, $\pmodel$,
         \item a stochastic physics-based model, $\amodel$, for the inadequacy of $\pmodel$ in describing $\dataset$,
    \end{enumerate}

    \noindent we seek to quantify the posterior probability density function of the combined set of unknown parameters of the two physical and inadequacy models $\paparam=\big\{\pparam,\aparam\big\}$.

\subsection{General Solution}
\label{sec:physInadNoisSolution}

    Our goal can be achieved by combining the approaches already developed in the previous sections \S\ref{sec:phys}, \S\ref{sec:physInad}, and \S\ref{sec:physNois}. First, note that the presence of model inadequacy requires us to use the modified form of \eqref{eq:physModel} as given in \eqref{eq:physInadResidual}. However, this equation takes as input, the truth, $\truth$, about the set of observations, $\dataset$. Since $\truthset$ is unknown, we have to consider all possibilities, $\ptruthset$, for $\truthset$, whose PDF is given by \eqref{eq:ptruthsetProbabilityJoint}. \\

    To do so, consider for the moment, a single realization, $\ptruthset$, of the truth dataset, $\truthset$, as defined in \eqref{eq:ptruthset}. The corresponding equations to \eqref{eq:physInadResidual}, \eqref{eq:inadRandset}, and \eqref{eq:physInadModel} for $\ptruthset$ would be then,
    \begin{eqnarray}
        \prandset &=& \ptruthset - \pmodel \big( \ptruthset, \pparam , \pscen \big) ~, \label{eq:physInadResidual*} \\
        \prandset &=& \{ \prand_1,\ldots,\prand_\ndo \} ~, \label{eq:inadRandset*} \\
        \prandset &\sim& \amodel \big( \ptruthset, \aparam, \pmodel ( \ptruthset, \pparam , \pscen ) \big) ~. \label{eq:physInadModel*}
    \end{eqnarray}

    Therefore, the modified set of equations corresponding to \eqref{eq:physInadLike}, \eqref{eq:physInadLikeJoint}, and \eqref{eq:physInadLikeIID} take the form,
    \begin{eqnarray}
        \like \big( \paparam &;& \ptruthset , \pamodel \big) ~\equiv~ \like \big( \paparam ~;~ \prandset , \pamodel \big) \label{eq:physInadLike*} \\
        &\equiv&     ~ \pi \big( \ptruthset \big| \paparam , \pamodel \big) \label{eq:physInadLikeJoint*} \\
        &\ceq{\iid}& ~ \prod_{i=1}^{\ndo} \pi \big( \ptruth_i \big| \aparam , \amodel \big( \ptruth_i, \aparam, \pmodel(\ptruth_i,\pparam,\pscen) \big) \big) \label{eq:physInadLikeIID*} ~.
    \end{eqnarray}

    Similar to \eqref{eq:physInadBayesRand} and \eqref{eq:physInadBayes}, the posterior probability density of $\paparam$ for a single realization, $\ptruthset$, of $\truthset$ can be then computed from the Bayes rule as,
    \begin{eqnarray}
        \label{eq:physInadBayes*}
        \pdfbig{\paparam \big| \ptruthset, \pamodel, \prior_\paparam} = \frac
        {
            \pdfbig{\ptruthset \big| \paparam , \pamodel} \pdfbig{\paparam \big| \pamodel,\prior_\paparam}
        }
        {
            \pdfbig{\ptruthset \big| \pamodel, \prior_\paparam}
        } ~,
    \end{eqnarray}

    However, the set $\ptruthset$ is only one possibility among the (potentially infinitely) many possible representations of the truth set, $\truthset$. Therefore, the likelihood in \eqref{eq:physInadLikeJoint*} has to be further modified to include not one, but all possibilities, $\ptruthset\in\ptruthSuperset$, for the reality, $\truthset$. \\

    Given the observed dataset, $\dataset$, and the associated set of the noise models, $\emodelset$, and their parameters, $\eparamset$, the probability of $\ptruthset$ being the truth, $\truthset$, is given by the posterior PDF in \eqref{eq:ptruthsetProbabilityJoint}. Thus, combining \eqref{eq:ptruthsetProbabilityJoint} with \eqref{eq:physInadLikeJoint*} yields the modified likelihood function of the model parameters as,
    \begin{eqnarray}
        \like \big( \paparam &;& \ptruthset , \dataset , \eparamset , \emodelset , \prior_\truthset , \pamodel \big) \nonumber \\
        &\equiv& ~ \pi \big( \ptruthset \big| \paparam , \pamodel \big)
        ~ \times ~ \pdfbig{ \ptruthset \big| \dataset, \eparamset, \emodelset, \prior_\truthset}
        \label{eq:physInadNoisLikeJoint*} \\
        &\ceq{\iid}& ~ \prod_{i=1}^{\ndo} \pi \big( \ptruth_i \big| \aparam , \amodel \big( \ptruth_i, \aparam, \pmodel(\ptruth_i,\pparam,\pscen) \big) \big) \nonumber \\
        &\times& ~ \pdfbig{ \ptruth_i \big| \data_i, \eparam[i], \emodel[i], \prior_{\truth_i}} ~.
        \label{eq:physInadNoisLikeIID*}
    \end{eqnarray}

    One can then marginalize \eqref{eq:physInadNoisLikeJoint*} over $\ptruthset$ to obtain the marginal likelihood of the parameters of the models, given only the {\it known quantities}: $\dataset$, $\eparamset$, $\emodelset$, $\prior_\truthset$, $\pmodel$, $\amodel$,
    \begin{eqnarray}
        \like \big( \paparam &;& \dataset , \eparamset , \emodelset , \prior_\truthset , \pamodel \big) \nonumber \\
        \label{eq:physInadNoisDataPDF}
        &\equiv& ~ \pi( \dataset , \eparamset, \emodelset | \paparam , \pamodel , \prior_\truthset ) \\
        \label{eq:physInadNoisLikeJoint}
        &\equiv& ~ \int_{\ptruthSuperset} ~ \pi \big( \ptruthset \big| \paparam , \pamodel \big)
        ~ \times ~ \pdfbig{ \ptruthset \big| \dataset, \eparamset, \emodelset, \prior_\truthset} ~ \diff\ptruthset \\
        &\ceq{\iid}& ~ \prod_{i=1}^{\ndo} ~ \int_{\sspace} \pi \big( \ptruth_i \big| \aparam , \amodel \big( \ptruth_i, \aparam, \pmodel(\ptruth_i,\pparam,\pscen) \big) \big) \nonumber \\
        \label{eq:physInadNoisLikeIID}
        &\times& ~ \pdfbig{ \ptruth_i \big| \data_i, \eparam[i], \emodel[i], \prior_{\truth_i}}
        ~ \diff\ptruth_i
        ~.
    \end{eqnarray}

\begin{figure}[t!]
        \begin{center}
            \includegraphics[width=\textwidth]{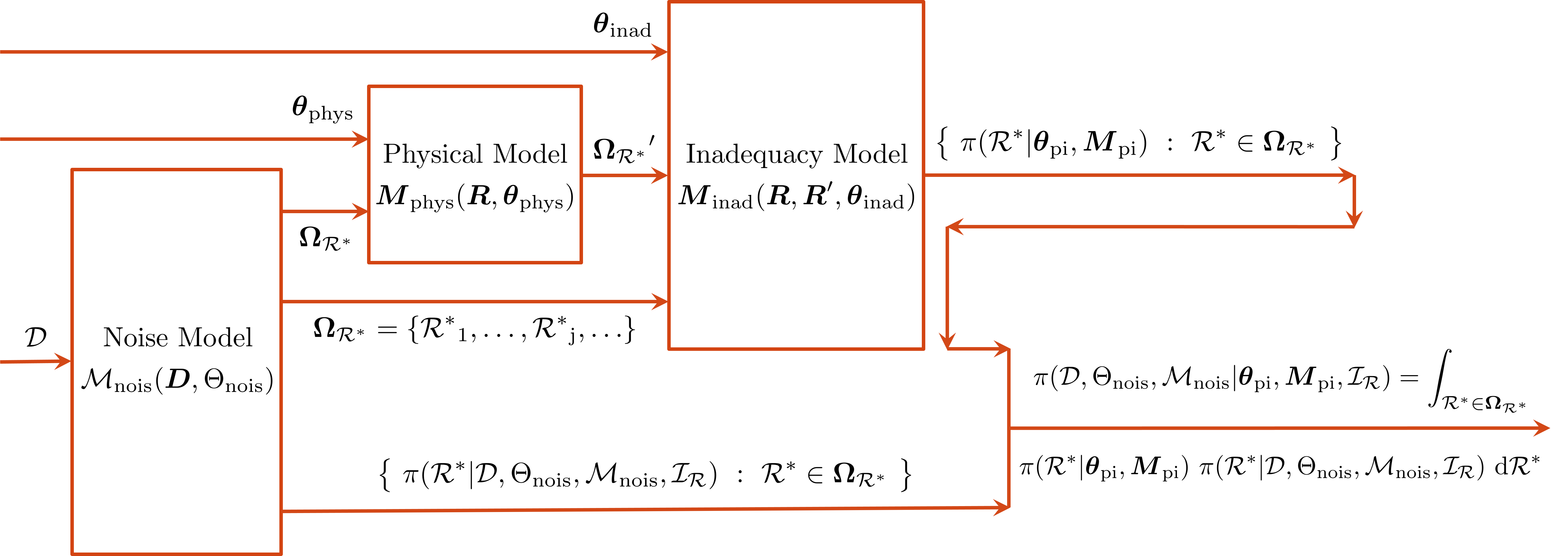}
        \end{center}
        \caption
        {
            A schematic illustration of the derivation of the likelihood function when the observational data, $\dataset$, is subject to measurement error described by the set of noise models, $\emodelset$, and their known parameters, $\eparamset$. In addition, the physical model, $\pmodel$, is inadequate, requiring an inadequacy model, $\amodel$, with unknown parameters, $\aparam$, whose values have to be constrained by data. Given $\dataTriplet$, there exists a range of possibilities, $\ptruthset$, for the unknown truth, $\truthset$. The set $\psspace$ represents the collection of all such possibilities. Then, each one of these possibilities is fed to the physical model, whose output, $\ptruthset'\in\psspace'$, is subsequently fed to the inadequacy model, $\amodel$, which in turn outputs the probability of obtaining $\ptruthset$ given $\ptruthset'$ for the specific set of values of the parameters, $\paparam=\{\pparam,\aparam\}$, of the physical and inadequacy models, $\pamodel=\{\pmodel,\amodel\}$. Simultaneously, the probability of $\ptruthset$ being $\truthset$ {\it given only} $\dataTriplet$ and the prior knowledge about the truth, $\prior_\truthset$, is also known. Therefore, the probability of $\ptruthset$ being $\truthset$ given all three physical, inadequacy, and noise models, as well as their parameter values, can be computed. Finally, the probability of $\dataTriplet$ being the correct representation of the truth, $\truthset$, given $\{\pamodel,\paparam,\prior_\truthset\}$, can be obtained by marginalizing the probability of $\ptruthset$ over all possibilities, $\ptruthset\in\psspace$, for the truth, $\truthset$.
        }
        \label{fig:physInadNoisLike}
    \end{figure}

    A schematic diagram illustrating the derivation of this likelihood function is given in Figure \ref{fig:physInadNoisLike}. Despite their similarities, there is a fine difference between the marginalization over $\ptruthset$ performed in the above likelihood function and the marginalization performed in the case of an ideal physical model in the presence of data uncertainty, which appears in \eqref{eq:physNoisBayes2}. The marginalization of \eqref{eq:physInadNoisLikeJoint*} and \eqref{eq:physInadNoisLikeIID*} is performed over all possible realizations of the truth, $\ptruthSuperset$, whether or not they satisfy \eqref{eq:physNoisModel}, whereas in \eqref{eq:physNoisParamLike} it is assumed that the physical model has no inadequacy in describing the truth. Therefore, the marginalization in \eqref{eq:physNoisParamLike} is strictly performed on a subset of $\ptruthSuperset$ that satisfy \eqref{eq:physNoisModel}. \\

    By contrast, not all possible realizations, $\ptruthset\in\ptruthSuperset$, of $\ptruthset$ in the hierarchical likelihood function of \eqref{eq:physInadNoisLikeJoint} have to necessarily satisfy \eqref{eq:physNoisModel}. For the construction of this likelihood function, we have already assumed that the physical model is inadequate in describing the truth, $\truthset$. Hence, the marginalization spans the entire sampling space of $\ptruthset$, which is $\ptruthSuperset$. \\

    Now, with the hierarchical likelihood function of $\paparam$ in hand, one can use the Bayes rule to write the posterior PDF of $\paparam$ as,
    \begin{eqnarray}
        \pi\big( \paparam &\big|& \dataset, \eparamset, \emodelset, \pamodel, \prior_\truthset, \prior_\paparam \big) \nonumber \\
        &=& \frac
        {
            \like \big( \paparam ~;~ \dataset, \eparamset, \emodelset, \prior_\truthset \big)
            ~
            \pdfbig{\paparam \big| \pamodel,\prior_\paparam}
        }
        {
            \pdfbig{ \dataset, \eparamset, \emodelset \big| \pamodel, \prior_\truthset, \prior_\paparam}
        }
        \label{eq:physInadNoisBayes1*} \\
        &=& \frac
        {
             \bigints_\psspace \pi \big( \ptruthset \big| \paparam , \pamodel \big)
            ~ \pdfbig{ \ptruthset \big| \dataset, \eparamset, \emodelset, \prior_\truthset} ~ \diff\ptruthset
            ~ \pdfbig{\paparam \big| \pamodel,\prior_\paparam}
        }
        {
            \pdfbig{ \dataset, \eparamset, \emodelset \big| \pamodel, \prior_\truthset, \prior_\paparam}
        }
        \label{eq:physInadNoisBayes2*} ~,
    \end{eqnarray}

    \noindent in which the denominator is again a factor that properly normalizes the posterior distribution to a posterior PDF. It gives the probability of all possibilities, $\ptruthSuperset$, for the truth, $\truthset$, where $\ptruthSuperset$ is fully determined by $\dataTriplet$,
    \begin{eqnarray}
        && \pdfbig{ \dataset, \eparamset, \emodelset \big| \pamodel, \prior_\truthset, \prior_\paparam} \equiv \pdfbig{ \ptruthSuperset \big| \pamodel, \prior_\truthset, \prior_\paparam} \label{eq:physInadNoisEvidence1} \\
        && = \int_\ppaspace \int_\sspace \pi \big( \ptruthset \big| \paparam , \pamodel \big)
           ~ \pdfbig{ \ptruthset \big| \dataset, \eparamset, \emodelset, \prior_\truthset}
           ~ \pdfbig{\paparam \big| \pamodel,\prior_\paparam}
           ~ \diff\ptruthset ~ \diff\paparam ~. \label{eq:physInadNoisEvidence2}
    \end{eqnarray}

    Plugging \eqref{eq:ptruthsetProbabilityJoint} into \eqref{eq:physInadNoisBayes2*} one gets,
    \begin{eqnarray}
        && \pi \big( \paparam \big| \dataset, \eparamset, \emodelset, \pamodel, \prior_\truthset, \prior_\paparam \big) \nonumber \\
        && = \frac
        {
            \bigints_\sspace
            ~ \pdfbig{ \dataset \big| \ptruthset, \eparamset, \emodelset }
            ~ \pdfbig{ \ptruthset \big| \prior_{\truthset} }
            ~ \pdfbig{ \ptruthset \big| \paparam , \pamodel }
            ~ \diff\ptruthset
            ~ \pdfbig{ \paparam \big| \pamodel, \prior_\paparam }
        }
        {
            \pdfbig{ \dataset \big| \eparamset, \emodelset, \prior_\truthset }
            ~ \pdfbig{ \dataset, \eparamset, \emodelset \big| \pamodel, \prior_\truthset, \prior_\paparam }
        }
        \label{eq:physInadNoisBayes3*} ~.
    \end{eqnarray}

    Thus, in a sense, the two latter terms, $\pdfbig{ \ptruthset \big| \prior_{\truthset} } ~ \pdfbig{ \ptruthset \big| \paparam , \pamodel }$, in the integrand of the numerator of \eqref{eq:physInadNoisBayes3*}, act like a prior probability on the likelihood of the observed dataset, $\pdfbig{ \dataset \big| \ptruthset, \eparamset, \emodelset }$, and correct its value according to the physical and inadequacy models in hand, for the given set of parameter values, $\paparam$. \\

    In the case of \iid~events, ignoring the normalization constants, the posterior PDF of \eqref{eq:physInadNoisBayes3*} takes the simple form,
    \begin{eqnarray}
        \pi \big( \paparam &\big|& \dataset, \eparamset, \emodelset, \pamodel, \prior_\truthset, \prior_\paparam \big) \nonumber \\
        &\propto& \pdfbig{ \paparam \big| \pamodel, \prior_\paparam } \nonumber \\
        &\times&
        \prod_{i=1}^\ndo \int_\sspace
        \pdfbig{ \data_i \big| \ptruth_i, \eparam[i], \emodel[i] }
        ~ \pdfbig{ \ptruth_i \big| \prior_{\truth_i} }
        ~ \pdfbig{ \ptruth_i \big| \paparam , \pamodel }
        ~ \diff\ptruth_i
        ~.
        \label{eq:physInadNoisBayes3IID*} 
    \end{eqnarray}

    Equations \eqref{eq:physInadNoisBayes3*} and \eqref{eq:physInadNoisBayes3IID*} describe the general form of the hierarchical (multilevel) posterior probability density function of the parameters of the physical model, $\pmodel$, whose inadequacy is described by the stochastic inadequacy model, $\amodel$, in the presence measurement error in data, whose effects are assumed to be fully determined by a set of noise models, $\emodelset$, and their known parameters, $\eparamset$. \qed

\vspace{0.2in}
\section*{Acknowledgement}
I thank J. Tinsley Oden, Ivo Babuska, and Fatemeh Bagheri for their helpful comments and valuable insights into many aspects of this manuscript.

\normalsize
\bibliographystyle{spbasic} 
\bibliography{./main} 

\end{document}